\documentclass[aps,rmp,square]{revtex4}
\usepackage{amsmath,amssymb}
\usepackage{graphicx}
\usepackage{pst-plot}
\usepackage{mathrsfs}
\usepackage{mathtools}
\usepackage{pst-func}
\usepackage{rotating}
\usepackage{epsfig}

\usepackage{chemarrow}

\usepackage[titletoc]{appendix}

\newtheorem{Joy}{Definition}[section]

\def\ocite#1{[\citenum{#1}]}

\setcitestyle{numbers}

\begin{document}

\title{${\,}$Whither All the Scope and Generality of Bell's Theorem?}


\author{Joy Christian}

\email{joy.christian@wolfson.ox.ac.uk}

\affiliation{Wolfson College, University of Oxford, Oxford OX2 6UD, United Kingdom}

\affiliation{Department of Materials, University of Oxford, Parks Road, Oxford OX1 3PH, England, UK}

\begin{abstract}
In a recent preprint James Owen Weatherall has attempted a simple local-deterministic model for the EPR-Bohm correlation
and speculated about why his model fails when my counterexample to Bell's theorem succeeds. Here I bring out the physical,
mathematical, and conceptual reasons why his model fails. In particular, I demonstrate why no model based on a tensor
representation of the rotation group SU(2) can reproduce the EPR-Bohm correlation. I demonstrate this by calculating
the correlation explicitly between measurement results ${{\mathscr A}=\pm\,1}$ and ${{\mathscr B}=\pm\,1}$ in a local
and deterministic model respecting the spinor representation of SU(2). I conclude by showing how Weatherall's reading of
my model is misguided, and bring out a number of misconceptions and unwarranted assumptions in his imitation of my
model as it relates to the Bell-CHSH inequalities.
\end{abstract}

\maketitle

\parskip 5pt

\section{Introduction}

In a recent preprint \ocite{Jim} James Owen Weatherall has attempted a local and deterministic model for the EPR-Bohm experiment,
which he affirms to have been inspired by my work on Bell's theorem
\ocite{disproof}\ocite{Christian}\ocite{Chris-refutation}\ocite{Chris-gill}\ocite{Chris-macro}.
Although his analysis mainly concerns his attempted reconstruction of my model and why the attempt
fails, his preprint has been worded in a
manner that has allowed some readers to embrace his discussion as a criticism of my work on Bell's theorem
\ocite{disproof}\ocite{Christian}\ocite{Chris-macro}. Here I demonstrate that (as he himself stresses to some extent)
the analysis Weatherall presents in his preprint has nothing to do with my model, or with the physics and
mathematics of the EPR-Bohm correlation \ocite{Chris-macro}\ocite{what}.
In fact his analysis exhibits no understanding of how my local-realistic framework works, nor of the
reasons why it explains the origins of all quantum correlations \ocite{Christian}\ocite{origins}. I show this
by first calculating the EPR-Bohm correlation in a successful local-deterministic
model based on the spinor representation of SU(2), and then revealing a number of misconceptions and
unwarranted assumptions in Weatherall's reconstruction of my model as it relates to the Bell-CHSH inequalities. 
I conclude that, contrary to first impressions, Weatherall's thinly veiled criticism of my work is entirely vacuous.

\section{An Exact Local-deterministic Model for the EPR-Bohm correlation}

In order to bring out the erroneous assumptions in Weatherall's analysis, it would be convenient to assess it in the
light of a successful model for the EPR-Bohm correlation. This will allow us to unveil his assumptions more easily. 

\subsection{A Complete Specification of the Singlet State}

To this end, let us recall Bell's definition of a {\it complete} theory \ocite{Bell}. He considered a physical theory to be
complete just in case its predictions for the EPR-Bohm experiment are dictated by local-deterministic functions of the form
\begin{equation}
{\mathscr A}({\bf r},\,\lambda): {\rm I\!R}^3\!\times\Lambda\longrightarrow S^0\equiv \{-1,\,+1\}, \label{map-1-1}
\end{equation}
where ${{\rm I\!R}^3}$ is the space of 3-vectors ${\bf r}$, ${\Lambda}$ is a space of the complete states ${\lambda}$, and
${S^0\equiv \{-1,\,+1\}}$ is a unit 0-sphere. He then proved a mathematical theorem concluding that no pair of functions of this
form can reproduce the correlation as strong as that
predicted by quantum mechanics for the rotationally invariant singlet state:
\begin{equation}
{\cal E}({\bf a},\,{\bf b})\,=\lim_{\,n\,\gg\,1}\left[\frac{1}{n}\sum_{k\,=\,1}^{n}\,
{\mathscr A}({\bf a},\,{\lambda}^k)\;{\mathscr B}({\bf b},\,{\lambda}^k)\right]
\,\not=\,-\,{\bf a}\cdot{\bf b}\,. \label{corre-1-1}
\end{equation}

As it stands, this conclusion of Bell is entirely correct and beyond dispute, provided we accept prescription (\ref{map-1-1})
as codifying a complete specification of the singlet state. In his preprint, following Bell and CHSH \ocite{Bell}\ocite{Clauser},
Weatherall accepts (\ref{map-1-1}) as codifying a complete specification of the singlet state, whereas my work begins by
recognizing that (\ref{map-1-1}) {\it does not}, and {\it cannot}, codify a complete specification of the singlet state
\ocite{Christian}\ocite{origins}. I have argued that Bell's prescription is based on an incorrect underpinning of both the
EPR argument \ocite{EPR} and the actual topological configurations involved in the EPR-Bohm experiments \ocite{Christian},
even if we leave the physics and mathematics underlying the correlation aside. My argument is rather subtle and requires
a clear understanding of what is meant by both a {\it function} in mathematics and the geometry and topology of a
parallelized 3-sphere. But the bottom line of the argument is that, for any two-level system, the EPR criterion of
completeness demands the correct measurement functions to be necessarily of the form
\begin{equation}
\pm\,1\,=\,{\mathscr A}({\bf r},\,\lambda):
{\rm I\!R}^3\!\times\Lambda\longrightarrow S^3 \sim {\rm SU(2)}, \label{map-1-2}
\end{equation}
with the {\it simply-connected} codomain ${S^3}$ of ${{\mathscr A}({\bf r},\,\lambda)}$ replacing the {\it totally disconnected}
codomain ${S^0}$ assumed by Bell. Thus ${{\mathscr A}({\bf r},\,\lambda)=\pm\,1}$ now represents a point of a parallelized 3-sphere,
${S^3}$. As a function, it takes values from the domain ${{\rm I\!R}^3\!\times\Lambda}$ and ends up {\it belonging} to the codomain
${S^3}$. Consequently, any correlation between a pair of such results is a correlation between points of a parallelized
3-sphere. Unless based on a prescription of this precise form, any Bell-type
analysis simply does not get off the ground, because without completeness there can be no theorem \ocite{Christian}.

Here ${S^3}$---which can be thought of as the
configuration space of all possible rotations of a rotating body (including spinorial sign changes)---is defined
as the set of all unit quaternions isomorphic to a unit parallelized 3-sphere:
\begin{equation}
S^3:=\left\{\,{\bf q}(\psi,\,{\bf r}):=\exp{\left[\,{\boldsymbol\beta}({\bf r})\,\frac{\psi}{2}\,\right]}\;
\Bigg|\;||\,{\bf q}(\psi,\,{\bf r})\,||^2=1\right\}\!, \label{nonoonpara}
\end{equation}
where ${{\boldsymbol\beta}({\bf r})}$ is a bivector rotating about ${{\bf r}\in{\rm I\!R}^3}$
with the rotation angle ${\psi}$ in the range ${0\leq\psi < 4\pi}$. Throughout this paper I will follow
the concepts, notations, and terminology of geometric algebra \ocite{Clifford}\ocite{Hestenes}. Accordingly,
${{\boldsymbol\beta}({\bf r})\in S^2\subset S^3}$ can be parameterized by a unit vector
${{\bf r}=r_1\,{\bf e}_1+r_2\,{\bf e}_2+r_3\,{\bf e}_3\in{\rm I\!R}^3}$ as
\begin{align}
{\boldsymbol\beta}({\bf r})\,&:=\,(\,I\cdot{\bf r}\,)\, \notag \\
&=\,r_1\,(\,I\cdot{\bf e}_1\,)
\,+\,r_2\,(\,I\cdot{\bf e}_2\,)\,+\,r_3\,(\,I\cdot{\bf e}_3\,) \notag \\
&=\,r_1\;{{\bf e}_2}\,\wedge\,{{\bf e}_3}
\,+\,r_2\;{{\bf e}_3}\,\wedge\,{{\bf e}_1}\,+\,r_3\;{{\bf e}_1}\,\wedge\,{{\bf e}_2}\,, 
\end{align}
with ${{\boldsymbol\beta}^2({\bf r})=-1}$. Here the trivector ${I:={\bf e}_1\,\wedge\,{\bf e}_2\,\wedge\,{\bf e}_3}$
(which also squares to ${-1}$) represents a volume form of the physical space \ocite{Clifford}\ocite{Hestenes}.
Each configuration of the rotating body can thus be represented by a quaternion of the form
\begin{equation}
{\bf q}(\psi,\,{\bf r})\,=\,\cos\frac{\psi}{2}\,+\,{\boldsymbol\beta}({\bf r})\,\sin\frac{\psi}{2}\,, \label{defi-2}
\end{equation}
which in turn can always be decomposed as a product of two bivectors belonging to an ${S^2\subset S^3\sim{\rm SU(2)}}$,
\begin{equation}
{\boldsymbol\beta}({\bf r''})\,{\boldsymbol\beta}({\bf r'})\,
=\,\cos\frac{\psi}{2}\,+\,{\boldsymbol\beta}({\bf r})\,\sin\frac{\psi}{2}\,, \label{defi-2-a}
\end{equation}
with ${\psi}$ being its rotation angle from ${{\bf q}(0,\,{\bf r})=1}$. Note also that ${{\bf q}(\psi,\,{\bf r})}$ reduces
to ${\pm\,1}$ as ${\psi\rightarrow\,2\kappa\pi}$ for ${\kappa\,=\,0,\,1,\;\text{or}\,\;2}$.

It is of paramount importance to note here that our topologically corrected prescription (\ref{map-1-2}) does not alter the
actual measurement results. 
For a given vector ${\bf r}$ and an initial state ${\lambda}$, both operationally and mathematically we still have
\begin{equation}
{\mathscr A}({\bf r},\,\lambda)\,=\,+\,1\;\,{\rm or}\;-1
\end{equation}
as the {\it image points} of the function ${{\mathscr A}({\bf r},\,\lambda)}$ as demanded by Bell, but now the topology of its
{\it codomain} has changed from a 0-sphere to a 3-sphere, with the latter embedded in ${{\rm I\!R}^4}$ in such a manner that the
prescriptions (\ref{map-1-1}) and (\ref{map-1-2}) are {\it operationally identical}
\ocite{Christian}\ocite{origins}. On the other hand, without
this topological correction it is impossible to provide a complete account of all possible measurement results in the sense
specified by EPR. Thus the selection of the codomain ${S^3\hookrightarrow {\rm I\!R}^4}$ in prescription (\ref{map-1-2}) is
not a matter of choice but necessity. What is responsible for the EPR correlation is {\it the topology of the set of all possible
measurement results}. But once the codomain of ${{\mathscr A}({\bf r},\,\lambda)}$ is so corrected, the proof of Bell's theorem
(as given in Refs.${\,}$\ocite{Bell}) simply falls apart. Moreover, it turns out that the strength of the correlation for
{\it any}${\,}$ physical system is entirely determined by the torsion within the codomain of the local functions
${{\mathscr A}({\bf r},\,\lambda)}$.

Returning to Weatherall's analysis, it should be clear now that, because it is based on a prescription other than (\ref{map-1-2}),
it too is a {\it non-starter} \ocite{Christian}. More importantly, once we recognize that the {\it only} way of providing a
complete account of all possible measurement results for the singlet state is by means of prescription (\ref{map-1-2}),
the statistical procedure for analyzing the correlation must be consistently customized for the
set ${S^3}$ of all unit quaternions,
{\it which Weatherall fails to do}.\break Since this procedure can be appreciated more readily by studying the
explicit construction of my model from my book and elsewhere, I now proceed to reproduce the model in some detail
in the following subsections. Doing so will also dispel a persistent but gravely disingenuous charge that my model encounters
``certain technical complications'' \ocite{Chris-refutation}\ocite{Chris-gill}.

\subsection{Construction of the Measurement Functions}

Once the measurement results are represented by functions of the form (\ref{map-1-2}), it is easy to reproduce the
EPR-Bohm correlation
in a manifestly local, realistic, and deterministic manner. This is because a parallelized 3-sphere has quite a unique and
distinctive topological structure \ocite{Chris-macro}\ocite{what}. 
It is one of the only two parallelizable spheres with non-vanishing
torsion---the other one, with variable torsion, being the 7-sphere. Once parallelized by a constant torsion, the 3-sphere
remains closed under multiplication, and forms one of the only four possible normed division algebras
\ocite{what}\ocite{origins}. These are profound concepts underlying the very existence and strength of
quantum correlations \ocite{Christian}. By ignoring them and dismissing them as irrelevant, Weatherall is ignoring
the physics and mathematics of the quantum correlations. More importantly, because of the
unique and distinctive topological characteristics of the 3-sphere \ocite{what}, the measurement functions such as
${{\mathscr A}({\bf r},\,\lambda)}$ for the EPR-Bohm correlation have to be constructed in a very specific manner for
any model to be successful \ocite{Christian}.
To this end, let us begin with the following definition of the orientation of a vector space:

\begin{Joy}
An orientation of a finite dimensional vector space ${{\cal V}_d}$ is an equivalence class of
ordered basis, say ${\{f_1,\,\dots,\,f_d\}}$, which determines the same orientation of ${{\cal V}_d}$ as the basis
${\{f'_1,\,\dots,\,f'_d\}}$ if ${f'_i =  \omega_{ij} f_j}$ holds with ${det(\omega_{ij})>0}$, and the opposite orientation
of ${{\cal V}_d}$ as the basis ${\{f'_1,\,\dots,\,f'_d\}}$ if ${f'_i = \omega_{ij} f_j}$ holds with ${det(\omega_{ij}) < 0}$.
\end{Joy}
(Here repeated indices are summed over.)
Thus each positive dimensional real vector space has precisely two possible orientations, which (rather suggestively)
can be denoted as ${\lambda=+1}$ or ${\lambda=-1}$. More generally an oriented smooth manifold such as ${S^3}$
consists of that manifold together with a choice of orientation for each of its tangent spaces.

It is important to note that orientation of a manifold is a {\it relative} concept \ocite{Milnor}.
In particular, the orientation of a tangent space ${{\cal V}_d}$ of a manifold defined by the equivalence class of
ordered basis such as ${\{f_1,\,\dots,\,f_d\}}$ is meaningful only with respect to that defined by the equivalence
class of ordered basis ${\{f'_1,\,\dots,\,f'_d\}}$, and vice versa. To be sure, we can certainly orient a manifold
absolutely by choosing a set of ordered bases for all of its tangent spaces, but the resulting manifold can be said
to be left or right oriented only with respect of another such set of ordered basis \ocite{Milnor}.

Now the natural configuration space for an EPR-Bohm type experiment is a unit parallelized 3-sphere, which can
be embedded in ${{\rm I\!R}^4}$ with a choice of orientation, say ${\lambda=+1}$ or ${-1}$. This choice of orientation can be
identified with the initial state of the particle pair in the singlet state
with respect to the orientation of the detector basis as follows.
We first characterize the embedding space ${{\rm I\!R}^4}$ by the graded basis
\begin{equation}
\left\{\,1,\;L_{1}(\lambda),\;L_{2}(\lambda),\;L_{3}(\lambda)\,\right\}, \label{sinbas}
\end{equation}
with ${\lambda = \pm\,1}$ representing the two possible orientations of ${S^3}$ and the basis elements
${L_{\mu}(\lambda)}$ satisfying the algebra
\begin{equation}
L_{\mu}(\lambda)\,L_{\nu}(\lambda) \,=\,-\,g_{\mu\nu}\,-\,\epsilon_{\mu\nu\rho}\,L_{\rho}(\lambda)\,, \label{wh-o88}
\end{equation}
with an arbitrary metric ${g_{{\mu}{\nu}}}$ on ${S^3}$.
Here the bivectors ${\{\,a_{\mu}\;L_{\mu}(\lambda)\,\}}$ will represent the spin angular
momenta of the particles, with ${\mu}$ = 1, 2, 3 and the repeated indices summed over.
These momenta can be assumed to be detected by the detector bivectors, say ${\{\,a_{\mu}\,D_{\mu}\,\}}$, with
the corresponding detector basis ${\left\{\,1,\,D_1,\,D_2,\,D_3\,\right\}}$ satisfying the algebra
\begin{equation}
D_{\mu}\,D_{\nu} \,=\,-\,g_{{\mu}{\nu}}\,-\,\epsilon_{{\mu}{\nu}{\rho}}\,D_{\rho} \label{wh-o99}
\end{equation}
and related to the spin basis ${\left\{\,1,\,L_{1}(\lambda),\,L_{2}(\lambda),\,L_{3}(\lambda)\,\right\}\;}$as
\begin{equation}
\left(\begin{array}{c} 1 \\ L_1(\lambda) \\ L_2(\lambda) \\ L_3(\lambda) \end{array}\right)\;=\;
\left(\begin{matrix} \; 1 \; & \; 0 \; & \; 0 \; & \; 0 \; \\
                     \; 0 \; & \; \lambda \; & \; 0 \; & \; 0 \; \\
                     \; 0 \; & \; 0 \; & \; \lambda \; & \; 0 \; \\
                     \; 0 \; & \; 0 \; & \; 0 \; & \; \lambda \;
\end{matrix}\right)
\left(\begin{array}{c} 1 \\ D_1 \\ D_2 \\ D_3 \end{array}\right)\!.
\end{equation}
Evidently, the determinant of this matrix works out to be ${det(\omega_{ij})= \lambda}$.
Since ${\lambda^2=+1}$ and ${\omega^2}$ is a ${4\times 4}$ identity matrix, this relation can be more succinctly written as
\begin{equation}
L_{\mu}(\lambda)\,=\,\lambda\,D_{\mu}\;\;\;\;\text{and}\;\;\;\;
D_{\mu}\,=\,\lambda\,L_{\mu}(\lambda)\,,\label{1237} 
\end{equation}
or equivalently as
\begin{equation}
\left\{\,1,\;L_{1}(\lambda),\;L_{2}(\lambda),\;L_{3}(\lambda)\,\right\}\,=\,
\left\{1,\,\lambda\,D_1,\,\lambda\,D_2,\,\lambda\,D_3\right\}\; \label{389-2}
\end{equation}
and
\begin{equation}
\left\{\,1,\;D_{1},\;D_{2},\;D_{3}\,\right\}\,=\,
\left\{1,\,\lambda\,L_1(\lambda),\,\lambda\,L_2(\lambda),\,\lambda\,L_3(\lambda)\right\}. \label{389-1}
\end{equation}
These relations reiterate the fact that orientation of any manifold is a {\it relative} concept. In particular, orientation of
${S^3}$ defined by the spin basis ${\{\,1,\,L_{\mu}(\lambda)\,\}}$ is meaningful only with respect to that defined by the detector
basis ${\{\,1,\,D_{\mu}\,\}}$ with the orientation ${\lambda=+1}$, and vice versa. Thus the spin basis are said to define the
{\it same} orientation of ${S^3}$ as the detector basis if ${L_{\mu}(\lambda=+1)=+D_{\mu}}$,
and the spin basis are said to define the {\it opposite}
orientation of ${S^3}$ as the detector basis if ${L_{\mu}(\lambda=-1)=-\,D_{\mu}}$.
Note also that the numbers ${1}$ and ${{\bf L}({\bf r},\,\lambda)}$ are treated here on equal footing.

We are now in a position to define the functions ${{\mathscr A}({\bf a},\,{\lambda})}$ and ${{\mathscr B}({\bf b},\,{\lambda})}$
as results of measurement interactions (or Clifford products) between detector bivectors ${-\,{\bf D}({\bf a})}$ and
${+\,{\bf D}({\bf b})}$ and spin bivectors ${{\bf L}({\bf a},\,\lambda)}$ and ${{\bf L}({\bf b},\,\lambda)}$ as follows:
\begin{align}
{\rm SU(2)}\sim S^3\ni\pm\,1\,=\,{\mathscr A}({\bf a},\,{\lambda^k})\,=\,\,-\,{\bf D}({\bf a})\,{\bf L}({\bf a},\,\lambda^k)
\,=\,\{-\,a_{\mu}\;D_{\mu}\,\}\,\{\,a_{\nu}\;L_{\nu}(\lambda^k)\,\}\,=\,
\begin{cases}
+\,1\;\;\;\;\;{\rm if} &\lambda^k\,=\,+\,1 \\
-\,1\;\;\;\;\;{\rm if} &\lambda^k\,=\,-\,1
\end{cases} \label{88-oi}
\end{align}
and
\begin{align}
{\rm SU(2)}\sim S^3\ni\pm\,1\,=\,{\mathscr B}({\bf b},\,{\lambda^k})\,=\,\,+\,{\bf D}({\bf b})\,{\bf L}({\bf b},\,\lambda^k)
\,=\,\{+\,b_{\mu}\;D_{\mu}\,\}\,\{\,b_{\nu}\;L_{\nu}(\lambda^k)\,\}\,=\,
\begin{cases}
-\,1\;\;\;\;\;{\rm if} &\lambda^k\,=\,+\,1 \\
+\,1\;\;\;\;\;{\rm if} &\lambda^k\,=\,-\,1\,,
\end{cases} \label{99-oi}
\end{align}
where the relative orientation ${\lambda}$ is now assumed to be a random variable, with 50/50 chance of being ${+1}$ or ${-\,1}$
at the moment of creation of the singlet pair of spinning particles. In what follows, I will
assume that the orientation of ${S^3}$ defined by
the detector basis ${\{\,1,\,D_{\nu}\,\}}$ has been fixed before hand \ocite{Christian}\ocite{Chris-macro}. Thus the spin bivector
${\{\,a_{\mu}\;L_{\mu}(\lambda)\,\}}$ is a random bivector with its handedness determined
{\it relative} to the detector bivector ${\{\,a_{\nu}\;D_{\nu}\,\}}$, by the relation
\begin{equation}
{\bf L}({\bf a},\,\lambda)
\,\equiv\,\{\,a_{\mu}\;L_{\mu}(\lambda)\,\}\,=\,\lambda\,\{\,a_{\nu}\;D_{\nu}\,\}\,\equiv\,\lambda\,{\bf D}({\bf a}), \label{OJS}
\end{equation}
where, as a direct consequence of the algebra (\ref{wh-o88}) with ${g_{{\mu}{\nu}}=\delta_{{\mu}{\nu}}}$,
the bivectors ${{\bf L}({\bf a},\,\lambda)}$ satisfy the following identity:
\begin{equation}
{\bf L}({\bf a},\,\lambda)\,{\bf L}({\bf a'},\,\lambda)\,=\,-\,{\bf a}\cdot{\bf a'}\,-\,
{\bf L}({\bf a}\times{\bf a'},\,\lambda).
\end{equation}
Using these relations the spin detection events (\ref{88-oi}) and (\ref{99-oi})
follow at once from the algebras defined in (\ref{wh-o88}) and (\ref{wh-o99}).

Evidently, the measurement results ${{\mathscr A}({\bf a},\,{\lambda})}$ and ${{\mathscr B}({\bf b},\,{\lambda})}$ as defined
above, in addition to being manifestly realistic, are strictly local and deterministically determined numbers. In fact,
they are not even contextual. Alice's measurement result ${{\mathscr A}({\bf a},\,{\lambda})}$---although it refers to a freely
chosen direction ${\bf a}$---depends only on the initial state ${\lambda}$; and likewise, Bob's measurement result
${{\mathscr B}({\bf b},\,{\lambda})}$---although it refers to a freely chosen direction ${\bf b}$---depends only on
the initial state ${\lambda}$. Let us also not overlook the fact that, as binary numbers,
${{\mathscr A}({\bf a},\,{\lambda})=\pm\,1}$ and ${{\mathscr B}({\bf b},\,{\lambda})=\pm\,1}$ are still points of a parallelized 
3-sphere. To confirm this, recall that a parallelized 3-sphere is a set of unit quaternions of the form 
\begin{equation}
{\bf q}^k(\psi,\,{\bf r},\,\lambda)\,:=\,\left\{\lambda^k\,\cos\frac{\psi}{2}\,+\,{\bf L}\!\left({\bf r},\,\lambda^k\right)
\,\sin\frac{\psi}{2}\,\right\}, \label{quat-0}
\end{equation}
and a measurement result such as ${{\mathscr A}({\bf a},\,\lambda)=\pm1}$ is a limiting case of such a quaternion constituting
the 3-sphere:
\begin{align}
S^3\ni\pm\,1\,=\,{\mathscr A}({\bf a},\,\lambda)\,
&=\,\lim_{{\bf a'}\rightarrow\,{\bf a}}\,{\mathscr A}({\bf a},\,{\bf a'},\,\lambda)\, \notag \\
&=\,\lim_{{\bf a'}\rightarrow\,{\bf a}}\,\left\{\,-\,{\bf D}({\bf a})\,{\bf L}({\bf a'},\,\lambda)\,\right\}\, \notag \\
&=\,\lim_{{\bf a'}\rightarrow\,{\bf a}}\,\left\{\,(\,-\,I\cdot{\bf a})(\,\lambda\,I\cdot{\bf a'})\,\right\} \notag \\
&=\,\lim_{{\bf a'}\rightarrow\,{\bf a}}\,\left\{\,\lambda\,{\bf a}\cdot{\bf a'}
                                   \,+\,\lambda\,I\cdot({\bf a}\times{\bf a'})\,\right\} \notag \\
&=\,\lim_{{\psi}\rightarrow\,{2\kappa\pi}}\,
                     \left\{\lambda\,\cos\frac{\psi}{2}\,+\,{\bf L}({\bf c},\,\lambda)\,\sin\frac{\psi}{2}\,\right\} \notag \\
&=\,\lim_{{\psi}\rightarrow\,{2\kappa\pi}}\,\left\{\,{\bf q}(\psi,\,{\bf c},\,\lambda)\,\right\}. \label{amanda}
\end{align}
Here ${I={\bf e}_1\,\wedge\,{\bf e}_2\,\wedge\,{\bf e}_3}$ is the volume form,
limit ${{\bf a'}\rightarrow\,{\bf a}}$ is equivalent to the limit ${\psi\rightarrow\,2\kappa\pi}$ for
${\kappa\,=\,0,\,1,\;\text{or}\,\;2}$, ${\psi=2\,\eta_{{\bf a}\,{\bf a'}}}$ is the rotation angle about
the axis ${{\bf c}:={\bf a}\times{\bf a'}/|{\bf a}\times{\bf a'}|,\,}$ and
${\eta_{{\bf a}\,{\bf a'}}}$ is the angle between ${\bf a}$ and ${\bf a'}$ \ocite{Chris-macro}.

\subsection{A Crucial Lesson from Basic Statistics}

It is important to note that the variables ${{\mathscr A}({\bf a},\,{\lambda})}$ and ${{\mathscr B}({\bf b},\,{\lambda})}$
defined in equations (\ref{88-oi}) and (\ref{99-oi}) are generated with {\it different} bivectorial scales of dispersion (or
different standard deviations) for each measurement direction ${\bf a}$ and  ${\bf b}$. Consequently,
in statistical terms these variables are raw scores, as opposed to standard scores \ocite{scores-1}.
Recall that a standard score, z, indicates how
many standard deviations an observation or datum is above or below the mean. If ${\rm x}$ is a raw (or unnormalized) score
and ${\overline{\rm x}}$ is its mean value, then the standard (or normalized) score, ${{\rm z}({\rm x})}$, is defined by
\begin{equation}
{\rm z}({\rm x})\,=\,\frac{{\rm x}\,-\,{\overline{\rm x}}}{\sigma({\rm x})}\,,
\end{equation}
where ${\sigma({\rm x})}$ is the standard deviation of ${\rm x}$. A standard score thus represents the distance between
a raw score and population mean in the units of standard deviation, and
allows us to make comparisons of raw scores that come from very different sources \ocite{Christian}\ocite{scores-1}.
In other words, the mean value of the standard score itself is always zero, with standard deviation unity.
In terms of these concepts the correlation between raw scores ${\rm x}$ and ${\rm y}$ is defined as
\begin{align}
{\cal E}({\rm x},\,{\rm y})\,&=\;\frac{\,{\displaystyle\lim_{\,n\,\gg\,1}}\left[{\displaystyle\frac{1}{n}}\,
{\displaystyle\sum_{k\,=\,1}^{n}}\,({\rm x}^k\,-\,{\overline{\rm x}}\,)\;
({\rm y}^k\,-\,{\overline{\rm y}}\,)\right]}{\sigma({\rm x})\;\sigma({\rm y})} \label{co} \\
&=\,\lim_{\,n\,\gg\,1}\left[\frac{1}{n}
\sum_{k\,=\,1}^{n}\,{\rm z}({\rm x}^k)\;{\rm z}({\rm y}^k)\right]. \label{stan}
\end{align}
It is vital to appreciate that covariance by itself---{\it i.e.}, the numerator of equation (\ref{co}) by itself---does not
provide the correct measure of association between the raw scores, not the least because it depends on different units and scales
(or different scales of dispersion) that may have been used in the measurements of such scores. Therefore, to arrive at the correct
measure of association between the raw scores one must either use equation (\ref{co}), with the product of standard
deviations in the denominator, or use covariance of the standardized variables, as in equation (\ref{stan}).

Now, as discussed above, the random variables ${{\mathscr A}({\bf a},\,{\lambda})}$ and ${{\mathscr B}({\bf b},\,{\lambda})}$ are
products of two factors---one random and another non-random. Within the variable ${{\mathscr A}({\bf a},\,{\lambda})}$
the bivector ${{\bf L}({\bf a},\,\lambda)}$ is a random factor---a function of the orientation ${\lambda}$, whereas
the bivector ${-\,{\bf D}({\bf a})}$ is a non-random factor, independent of the orientation ${\lambda\,}$: 
\begin{align}
{\mathscr A}({\bf a},\,{\lambda})\,&=\,\,-\,{\bf D}({\bf a})\,{\bf L}({\bf a},\,\lambda) \\
\text{and}\;\;\;{\mathscr B}({\bf b},\,{\lambda})\,&=\,\,+\,{\bf D}({\bf b})\,{\bf L}({\bf b},\,\lambda)
\end{align}
Thus, as random variables, ${{\mathscr A}({\bf a},\,{\lambda})}$ and ${{\mathscr B}({\bf b},\,{\lambda})}$ are
generated with {\it different} standard deviations---{\it i.e.}, {\it different} sizes of the typical error.
More specifically, ${{\mathscr A}({\bf a},\,{\lambda})}$ is generated with the standard deviation
${-\,{\bf D}({\bf a})}$, whereas ${{\mathscr B}({\bf b},\,{\lambda})}$ is generated
with the standard deviation ${+\,{\bf D}({\bf b})}$. These two deviations
can be calculated as follows. Since errors in the linear relations propagate linearly,
the standard deviation ${\sigma({\mathscr A}\,)}$ of ${{\mathscr A}({\bf a},\,{\lambda})}$ is equal to ${-\,{\bf D}({\bf a})}$
times the standard deviation of ${{\bf L}({\bf a},\,\lambda)}$ [which I will denote as
${\sigma({A})=\sigma({\bf L}_{\bf a})}$], whereas the standard deviation ${\sigma({\mathscr B}\,)}$ of
${{\mathscr B}({\bf b},\,{\lambda})}$ is equal to ${+\,{\bf D}({\bf b})}$ times the standard deviation of
${{\bf L}({\bf b},\,\lambda)}$ [which I will denote as ${\sigma({B})=\sigma({\bf L}_{\bf b})}$]:
\begin{align}
\sigma({\mathscr A}\,)\,&=\,-\,{\bf D}({\bf a})\,\sigma({A}) \\
\text{and}\;\;\;\sigma({\mathscr B}\,)\,&=\,+\,{\bf D}({\bf b})\,\sigma({B}).
\end{align}
But since the bivector ${{\bf L}({\bf a},\,\lambda)}$ is normalized to unity, and since its mean
value ${m({\bf L}_{\bf a})}$ vanishes on the account of ${\lambda}$ being a
fair coin, its standard deviation is easy to calculate, and it turns out to be equal to unity:
\begin{equation}
\sigma({A})\,=\,\sqrt{\frac{1}{n}\sum_{k\,=\,1}^{n}\,\left|\left|\,A({\bf a},\,{\lambda}^k)\,-\,
{\overline{A({\bf a},\,{\lambda}^k)}}\;\right|\right|^2\,}\,
=\,\sqrt{\frac{1}{n}\sum_{k\,=\,1}^{n}\,
\left|\left|\,{\bf L}({\bf a},\,\lambda^k)\,-\,0\,\right|\right|^2\,}\,=\,1, \label{var-c}
\end{equation}
where the last equality follows from the normalization of ${{\bf L}({\bf a},\,\lambda)}$.
Similarly, it is easy to see that ${\sigma({B})}$ is also equal to ${1}$. Consequently, the standard deviation of
${{\mathscr A}({\bf a},\,{\lambda})=\pm\,1}$ works out to be ${-\,{\bf D}({\bf a})}$, and the standard
deviation of ${{\mathscr B}({\bf b},\,{\lambda})=\pm\,1}$ works out to be ${+\,{\bf D}({\bf b})}$. Putting
these two results together, we arrive at the following standard scores corresponding to the raw scores
${\mathscr A}$ and ${\mathscr B}$:
\begin{equation}
A({\bf a},\,{\lambda})=\frac{\,{\mathscr A}({\bf a},\,{\lambda})\,-\,
{\overline{{\mathscr A}({\bf a},\,{\lambda})}}}{\sigma({\mathscr A})}
\,=\,\frac{\,-\,{\bf D}({\bf a})\,{\bf L}({\bf a},\,\lambda)\,-\,0\,}{-\,{\bf D}({\bf a})}
\,=\,{\bf L}({\bf a},\,\lambda) \label{nonvar-a}
\end{equation}
and
\begin{equation}
B({\bf b},\,{\lambda})=\frac{\,{\mathscr B}({\bf b},\,{\lambda})\,-\,
{\overline{{\mathscr B}({\bf b},\,{\lambda})}}}{\sigma({\mathscr B})}
\,=\,\frac{\,+\,{\bf D}({\bf b})\,{\bf L}({\bf b},\,\lambda)\,-\,0\,}{+\,{\bf D}({\bf b})}
\,=\,{\bf L}({\bf b},\,\lambda),\label{nonvar-b}
\end{equation}
where I have used identities such as ${-\,{\bf D}({\bf a}){\bf D}({\bf a})=+1}$. Needless to say, these standard scores are
pure bivectors:
\begin{align}
{\rm SU(2)}\sim S^3\supset S^2\ni {\bf L}({\bf a},\,\lambda)\,=\, \pm\,1\;\,{\rm about}\;{{\bf a}}\in{\rm I\!R}^3, \\
\text{and}\;\;\;{\rm SU(2)}\sim S^3\supset S^2\ni {\bf L}({\bf b},\,\lambda)
\,=\, \pm\,1\;\,{\rm about}\;{{\bf b}}\in{\rm I\!R}^3. 
\end{align}

\subsection{How Errors Propagate within a Parallelized 3-sphere}

As noted towards the end of subsection II${\,}$A, one of several oversights in Weatherall's reading of my model
concerns his failure to recognize the necessity of applying the correct statistical procedure for analyzing the correlation
between the measurement results defined by (\ref{88-oi}) and (\ref{99-oi}). This is surprising, because I pointed
out this oversight to him in a private correspondence more than eighteen months ago. As we noted above, a parallelized 3-sphere
is a set of unit quaternions. Each point of a parallelized 3-sphere is thus represented by a unit quaternion. As a result,
the correct statistical procedure within my model must take into account how errors propagate in a 3-sphere of unit
quaternions.

Accordingly, let a probability density function ${P({\bf q}):S^3\rightarrow\,[{\hspace{1pt}}0,\,1]}$ of random quaternions
over ${S^3}$ be defined as:
\begin{equation}
P({\bf q})\,=\,\frac{1}{\sqrt{2\pi\left|\left|{\hspace{1pt}}\sigma({\bf q})\right|\right|^2\,}\;}\,
\exp\left\{-\,\frac{\,\left|\left|\,{\bf q}-m({\bf q})\right|\right|^2}{2\,
\left|\left|{\hspace{1pt}}\sigma({\bf q})\right|\right|^2}\right\},\label{probdensi-8}
\end{equation}
where the square root of ${\bf q=p\,p}$, ${{\bf p}\in S^3}$, is defined as
\begin{equation}
\sqrt{\,{\bf q}\,}\,=\sqrt{\,{\bf p}\,{\bf p}\,}\,:=\,\pm\,{\bf p}^{\dagger}(\,{\bf p}\,{\bf p}\,)
\,=\,\pm(\,{\bf p}^{\dagger}{\bf p}\,)\,{\bf p}\,=\,\pm\,{\bf p}\,. \label{1root1}  
\end{equation}
It is a matter of indifference whether the distribution of ${{\bf q}\in S^3}$ so chosen happens to be normal or not. Here
${\bf q}$ is an arbitrary quaternion within ${S^3(\lambda)}$ of the form (\ref{quat-0}), which is a sum of a scalar and a
bivector (treated on equal footing), with ${0\leq\psi\leq 4\pi}$ being the double-covering rotation angle about ${\bf r}$-axis.
The mean value of ${\bf q}$ is defined as
\begin{equation}
m({\bf q})\,=\,\frac{1}{n}\sum_{k\,=\,1}^{n}\,{\bf q}^k\,,
\end{equation}
and the standard deviation of ${{\bf q}}$ is defined as
\begin{equation}
\sigma[{\bf q}(\psi,\,{\bf r},\,\lambda)]\,
:=\,\sqrt{\frac{1}{n}\sum_{k\,=\,1}^{n}\,\left\{\,{\bf q}^k(\psi)\,-\,m({\bf q})\right\}\,
\left\{\,{\bf q}^k(2\pi-\psi)\,-\,m({\bf q})\right\}^{\dagger}\;}.\label{standipo}
\end{equation}

Note that in this definition ${{\bf q}(\psi)}$ is
coordinated by ${\psi}$ to rotate from ${0}$ to ${2\pi}$, whereas the conjugate ${{\bf q}^{\dagger}(2\pi-\psi)}$ is
coordinated by ${\psi}$ to rotate from ${2\pi}$ to ${0}$. Thus, for a given value of ${\lambda}$, both ${{\bf q}(\psi)}$
and ${{\bf q}^{\dagger}(2\pi-\psi)}$ represent the same sense of rotation about ${\bf r}$ (either both represent clockwise
rotations or both represent counterclockwise rotations). This is crucial for the calculation of standard deviation, for
it is supposed to give the average rotational distance within ${S^3}$ from its mean, with the average being taken, not over
rotational distances within a fixed orientation of ${S^3}$, but over the changes in the orientation ${\lambda}$ of ${S^3}$
itself. Note also that, according to the definition (\ref{quat-0}),
${{\bf q}(\psi)}$ and its conjugate ${{\bf q}^{\dagger}(\psi)}$ satisfy the following relation:
\begin{equation}
{\bf q}^{\dagger}(2\pi-\psi)\,=\,-\,{\bf q}(\psi)\,. \label{relat}
\end{equation}
Consequently, the standard deviation of both ${{\bf q}^{\dagger}(2\pi-\psi)}$ and ${-\,{\bf q}(\psi)}$
must, with certainty, give the same number:
\begin{equation}
\sigma[\,{\bf q}^{\dagger}(2\pi-\psi)]\,\equiv\,\sigma[\,-\,{\bf q}(\psi)]\,.
\end{equation}
It is easy to verify that definition (\ref{standipo}) for the standard deviation of ${{\bf q}(\psi)}$ does indeed satisfy
this requirement, at least when ${m({\bf q})=0}$. What is more, from equation (\ref{relat}) we note that the quantity being
averaged in the definition (\ref{standipo}) is essentially ${-\,{\bf q}\,{\bf q}}$. This quantity is insensitive to spinorial
sign changes such as ${{\bf q}\rightarrow -\,{\bf q}}$, but transforms into the quantity
${-\,{\bf q}^{\dagger}{\bf q}^{\dagger}}$ under
orientation changes such as ${{\lambda}\rightarrow -{\lambda}}$. By contrast, the quantity
${-\,{\bf q}\,{\bf q}^{\dagger}}$ would be insensitive to both spinorial sign changes as well as orientation changes.
Thus ${\sigma[{\bf q}(\lambda)]}$, as defined in (\ref{standipo}), is designed to remain sensitive to orientation changes
for correctly computing its averaging function on ${{\bf q}(\lambda)}$ in the present context.

Now, in order to evaluate ${\sigma({\mathscr A})}$ and ${\sigma({\bf L}_{\bf a})}$, we can rewrite the quaternion (\ref{quat-0})
rotating about ${{\bf r}={\bf a}}$ as a product
\begin{equation}
{\bf q}(\psi,\,{\bf a},\,\lambda)\,=\,{\bf p}(\psi,\,{\bf a})\,{\bf L}({\bf a},\,\lambda) \label{pro}
\end{equation}
of a non-random, non-pure quaternion
\begin{equation}
{\bf p}(\psi,\,{\bf a})\,:=\,\sin\left(\frac{\psi}{2}\right)\,-\,{\bf D(a)}\,\cos\left(\frac{\psi}{2}\right)
\,=\,\exp\left\{-\,{\bf D(a)}\!\left(\frac{\pi-\psi}{2}\right)\!\right\}
\label{genno}
\end{equation}
and a random, unit bivector ${{\bf L}({\bf a},\,\lambda)}$ satisfying
\begin{equation}
\frac{1}{n}\sum_{k\,=\,1}^{n}\,
{\bf L}({\bf a},\,\lambda^k)\,{\bf L}^{\dagger}({\bf a},\,\lambda^k)\,=\,1\,. \label{normL}
\end{equation}
Note that ${{\bf p}(\psi,\,{\bf a})}$ reduces to the unit bivector ${\pm\,{\bf D}\!\left({\bf a}\right)}$ for rotation
angles ${\psi=0}$, ${\psi=2\pi}$, and ${\psi=4\pi}$. Moreover, using the relations
${{\bf L}\!\left({\bf a},\,\lambda\right)=\lambda\,{\bf D}\!\left({\bf a}\right)}$ and ${{\bf D}^2({\bf a})=-\,1}$
it can be easily checked that the product in
(\ref{pro}) is indeed equivalent to the quaternion defined in (\ref{quat-0}) for ${{\bf r}={\bf a}}$. It is also easy to check
that the non-random quaternion ${{\bf p}(\psi,\,{\bf a})}$ satisfies the following relation with its conjugate:
\begin{equation}
{\bf p}^{\dagger}(2\pi-\psi,\,{\bf a})\,=\,{\bf p}(\psi,\,{\bf a})\,.
\end{equation}
Consequently we have
\begin{equation}
{\bf q}^{\dagger}(2\pi-\psi,\,{\bf a},\,\lambda)\,=\,\left\{{\bf p}
(2\pi-\psi,\,{\bf a})\,{\bf L}({\bf a},\,\lambda)\right\}^{\dagger}
\,=\,{\bf L}^{\dagger}({\bf a},\,\lambda)\,{\bf p}^{\dagger}(2\pi-\psi,\,{\bf a})
\,=\,{\bf L}^{\dagger}({\bf a},\,\lambda)\,{\bf p}(\psi,\,{\bf a})\,. \label{non-pro}
\end{equation}
Thus, substituting for ${{\bf q}(\psi,\,{\bf a},\,\lambda)}$ and ${{\bf q}^{\dagger}(2\pi-\psi,\,{\bf a},\,\lambda)}$ from
Eqs.${\,}$(\ref{pro}) and (\ref{non-pro}) into Eq.${\,}$(\ref{standipo}), together with
\begin{equation}
m({\bf q}_{\bf a}^{\,})\,=\,\frac{1}{n}\sum_{k\,=\,1}^{n}\,{\bf q}_{\bf a}^k
\,=\,{\bf p}(\psi,\,{\bf a})\,\left\{\frac{1}{n}\sum_{k\,=\,1}^{n}\,{\bf L}({\bf a},\,\lambda^k)\right\}
\,=\,{\bf p}(\psi,\,{\bf a})\,\left\{\frac{1}{n}\sum_{k\,=\,1}^{n}\,\lambda^k\right\}\,{\bf D}({\bf a})
\,=\,0\,,
\end{equation}
we have
\begin{align}
\sigma[{\bf q}(\psi,\,{\bf a},\,\lambda)]\,
&=\sqrt{\frac{1}{n}\sum_{k\,=\,1}^{n}\,\left\{\,{\bf p}(\psi,\,{\bf a})\,{\bf L}({\bf a},\,\lambda^k)\,\right\}\,
\left\{\,{\bf L}^{\dagger}({\bf a},\,\lambda^k)\,{\bf p}(\psi,\,{\bf a})\,\right\}\,} \notag \\
&=\,\sqrt{\,{\bf p}(\psi,\,{\bf a})\,\left\{\,\frac{1}{n}\sum_{k\,=\,1}^{n}\,
{\bf L}({\bf a},\,\lambda^k)\,{\bf L}^{\dagger}({\bf a},\,\lambda^k)\,\right\}\,{\bf p}(\psi,\,{\bf a})\,} \notag \\
&=\,\sqrt{\,{\bf p}(\psi,\,{\bf a})\,{\bf p}(\psi,\,{\bf a})\,}\, \notag \\
&=\,\pm\,{\bf p}(\psi,\,{\bf a})\,. \label{immminot}
\end{align}
Here I have used the normalization of ${{\bf L}({\bf a},\,\lambda)}$ as in (\ref{normL}), and the last equality follows from
the definition (\ref{1root1}). It can also be deduced from the polar form of the product
\begin{equation}
{\bf p}(\psi,\,{\bf a})\,{\bf p}(\psi,\,{\bf a})\,
=\,\cos\left(\pi-\psi\right)\,-\,{\bf D(a)}\,\sin\left(\pi-\psi\right)
=\,\exp\left\{\,-\,{\bf D(a)}\left(\pi-\psi\right)\right\}\,.
\end{equation}

The result for the standard deviation we have arrived at, namely
\begin{equation}
\sigma[\,{\bf q}(\psi,\,{\bf a},\,\lambda)]\,=\,\pm\,{\bf p}(\psi,\,{\bf a})\,, \label{minot}
\end{equation}
is valid for all possible rotation angles ${\psi}$ between the detector bivector
${-\,{\bf D}({\bf a})}$ and the spin bivector ${{\bf L}({\bf a},\,\lambda)}$. For the special
cases when ${\psi=0}$, ${\pi}$, ${2\pi}$, ${3\pi}$, and  ${4\pi}$, it reduces to the following set of standard deviations:
\begin{align}
\sigma[\,{\bf q}(\psi=0,\,{\bf a},\,\lambda)]\,&=\,\sigma({\mathscr A})\,=\,\pm\,{\bf D}({\bf a}) \notag \\
\sigma[\,{\bf q}(\psi=\pi,\,{\bf a},\,\lambda)]\,&=\,\sigma({\bf L}_{\bf a})\,=\,\pm\,1 \notag \\
\sigma[\,{\bf q}(\psi=2\pi,\,{\bf a},\,\lambda)]\,&=\,\sigma({\mathscr A})\,=\,\pm\,{\bf D}({\bf a}) \notag \\
\sigma[\,{\bf q}(\psi=3\pi,\,{\bf a},\,\lambda)]\,&=\,\sigma({\bf L}_{\bf a})\,=\,\pm\,1 \notag \\
\text{and}\;\;\;\sigma[\,{\bf q}(\psi=4\pi,\,{\bf a},\,\lambda)]\,&=\,\sigma({\mathscr A})
\,=\,\pm\,{\bf D}({\bf a})\,. \label{setresto}
\end{align}
To understand the physical significance of these results, let us first consider the special case when ${\psi=\pi}$. Then
\begin{equation}
{\bf q}(\psi=\pi,\,{\bf a},\,\lambda)\,=\,+\,{\bf L}({\bf a},\,\lambda)\,,
\end{equation}
which can be seen as such from the definition (\ref{quat-0}) above. Similarly, for the conjugate of
${{\bf q}(\psi=\pi,\,{\bf a},\,\lambda)}$ we have
\begin{equation}
{\bf q}^{\dagger}(\psi=\pi,\,{\bf a},\,\lambda)\,=\,-\,{\bf L}({\bf a},\,\lambda)
\,=\,+\,{\bf L}^{\dagger}({\bf a},\,\lambda)\,.
\end{equation}
Moreover, we have ${m({\bf L}_{\bf a})=0}$,
since ${{\bf L}\!\left({\bf a},\,\lambda\right)=+\,\lambda\,{\bf D}\!\left({\bf a}\right)}$
with ${\lambda=\pm\,1}$ being a fair coin. Substituting these results into definition (\ref{standipo})---together
with ${\psi=\pi}$---we arrive at
\begin{equation}
\sigma({\bf L}_{\bf a})\,=\,\sqrt{\frac{1}{n}\sum_{k\,=\,1}^{n}\,
{\bf L}({\bf a},\,\lambda^k)\,{\bf L}^{\dagger}({\bf a},\,\lambda^k)\;}\,=\,\pm\,1\,,
\end{equation}
since ${{\bf L}({\bf a},\,\lambda)\,{\bf L}^{\dagger}({\bf a},\,\lambda)=1}$.
Similarly, we can consider the case when ${\psi=3\pi}$ and again arrive at ${\sigma({\bf L}_{\bf a})=\pm\,1}$.

Next, we consider the three remaining special cases, namely ${\psi=0}$, ${2\pi}$, or ${4\pi}$. These cases correspond to
the measurement results, as defined, for example, in equation (\ref{88-oi}). To confirm this, recall from equation
(\ref{amanda}) that a measurement
result such as ${{\mathscr A}({\bf a},\,\lambda)=\pm1}$ is a limiting case of the quaternion (\ref{quat-0}).
If we now rotate the bivector ${{\bf L}({\bf c},\,\lambda)}$ to ${{\bf L}({\bf a},\,\lambda)}$ as
\begin{equation}
{\bf D}({\bf r})\,{\bf L}({\bf c},\,\lambda)\,{\bf D}^{\dagger}({\bf r})\,=\,{\bf L}({\bf a},\,\lambda)
\end{equation}
using some ${{\bf D}({\bf r})}$, and multiply Eq.{\,}(\ref{amanda}) from the left by ${{\bf D}({\bf r})}$ and
from the right by ${{\bf D}^{\dagger}({\bf r})}$, then we arrive at
\begin{align}
\!\!\!{\mathscr A}({\bf a},\,\lambda)\,
&=\,\lim_{{\bf a'}\rightarrow\,{\bf a}}\,\left\{{\bf D}({\bf r})\,{\bf q}(\psi,\,{\bf c},\,\lambda)\,
{\bf D}^{\dagger}({\bf r})\right\} \notag \\
&=\,\lim_{{\psi}\rightarrow\,{2\kappa\pi}}\,
                     \left\{\lambda\,\cos\frac{\psi}{2}\,+\,{\bf L}({\bf a},\,\lambda)\,\sin\frac{\psi}{2}\,\right\} \notag \\
&=\,\lim_{{\psi}\rightarrow\,{2\kappa\pi}}\,\left\{\,{\bf q}(\psi,\,{\bf a},\,\lambda)\,\right\} \notag \\
&=\,\lim_{{\psi}\rightarrow\,{2\kappa\pi}}\,\left\{\,{\bf p}(\psi,\,{\bf a})\,{\bf L}({\bf a},\,\lambda)\,\right\}, \label{knox}
\end{align}
where ${{\bf p}(\psi,\,{\bf a})}$ is defined in equation (\ref{genno}).
The limit ${{\bf a'}\rightarrow\,{\bf a}}$ is thus physically equivalent to the limit
${\psi\rightarrow\,2\kappa\pi}$ for ${\kappa\,=\,0,\,1,\;\text{or}\,\;2}$.
We therefore have the following relation between 
${{\mathscr A}({\bf a},\,\lambda)}$ and ${{\bf q}(\psi,\,{\bf a},\,\lambda)}$:
\begin{equation}
{\bf q}(\psi=2\kappa\pi,\,{\bf a},\,\lambda)\,
=\,\pm\;{\bf D}({\bf a})\,{\bf L}({\bf a},\,\lambda)\,
=\,\pm\,{\mathscr A}({\bf a},\,\lambda)\,,
\end{equation}
and similarly between ${{\mathscr A}^{\dagger}({\bf a},\,\lambda)}$ and ${{\bf q}^{\dagger}(\psi,\,{\bf a},\,\lambda)}$:
\begin{equation}
{\bf q}^{\dagger}(\psi=2\kappa\pi,\,{\bf a},\,\lambda)
\,=\,\left\{\,\pm\,{\bf D}({\bf a})\,{\bf L}({\bf a},\,\lambda)\,\right\}^{\dagger}
\,=\,\pm\,{\mathscr A}^{\dagger}({\bf a},\,\lambda)\,. \label{15-2}
\end{equation}
For example, for ${\psi=0}$ the definition (\ref{quat-0}) leads to
\begin{equation}
{\bf q}(\psi=0,\,{\bf a},\,\lambda)\,=\,-\,{\bf D}({\bf a})\,{\bf L}({\bf a},\,\lambda)
\,=\,+\,{\mathscr A}({\bf a},\,\lambda)\,. \label{14}
\end{equation}
This tells us that in the ${\psi\rightarrow\,0}$ limit the quaternion ${{\bf q}(\psi,\,{\bf a},\,\lambda)}$ reduces to the
scalar point ${\,-\,{\bf D}({\bf a})\,{\bf L}({\bf a},\,\lambda)}$ of ${S^3}$. Moreover, we have ${m({\mathscr A})=0}$, since
${m({\bf L}_{\bf a})=0}$
as we saw above. On the other hand, from the definition (\ref{quat-0}) of ${{\bf q}(\psi,\,{\bf a},\,\lambda)}$
we also have the following relation between the
conjugate variables ${{\mathscr A}^{\dagger}({\bf a},\,\lambda)}$ and ${{\bf q}^{\dagger}(\psi=2\pi,\,{\bf a},\,\lambda)}$:
\begin{equation}
{\bf q}^{\dagger}(\psi=2\pi,\,{\bf a},\,\lambda)
\,=\,+\,\left\{\,{\bf D}({\bf a})\,{\bf L}({\bf a},\,\lambda)\,\right\}^{\dagger}
\,=\,+\,{\bf L}^{\dagger}({\bf a},\,\lambda)\,{\bf D}^{\dagger}({\bf a})
\,=\,-\,{\bf L}^{\dagger}({\bf a},\,\lambda)\,{\bf D}({\bf a})
\,=\,-\,{\mathscr A}^{\dagger}({\bf a},\,\lambda)\,. \label{15-1}
\end{equation}
This tells us that in the ${\psi\rightarrow\,2\,\pi}$ limit the quaternion ${{\bf q}^{\dagger}(\psi,\,{\bf a},\,\lambda)}$
reduces to the scalar point ${\,-\,{\bf L}^{\dagger}({\bf a},\,\lambda)\,{\bf D}({\bf a})}$ of ${S^3}$.
Thus the case ${\psi=0}$ does indeed correspond to the measurement events. The physical significance of the two remaining
cases, namely ${\psi=2\pi}$ and ${4\pi}$, can be verified similarly, confirming the set of results listed in (\ref{setresto}):
\begin{equation}
\sigma({\mathscr A})\,=\,\pm\,{\bf D}({\bf a})\,. \label{a30}
\end{equation}
Substituting this and ${\sigma({\mathscr B})\,=\,\pm\,{\bf D}({\bf b})}$
into Eqs.${\,}$(\ref{nonvar-a}) and (\ref{nonvar-b}) then immediately leads to the standard scores:
\begin{align}
{A}({\bf a},\,\lambda)
\,&=\frac{\,\pm\,{\mathscr A}({\bf a},\,{\lambda})\,-\,
{\overline{{\mathscr A}({\bf a},\,{\lambda})}}}{\sigma({\mathscr A})}
\,=\frac{\,\pm\;{\bf D}({\bf a})\,{\bf L}({\bf a},\,\lambda)\,-\,0\,}{\sigma({\mathscr A})}
\,=\,\left\{\frac{\,\pm\,{\bf D}({\bf a})\,}{\sigma({\mathscr A})}\right\}\,{\bf L}({\bf a},\,\lambda)
\,=\,{\bf L}({\bf a},\,\lambda) \label{standj}\\
\text{and}\;\;\;{B}({\bf b},\,\lambda)
\,&=\frac{\,\pm\,{\mathscr B}({\bf b},\,{\lambda})\,-\,
{\overline{{\mathscr B}({\bf b},\,{\lambda})}}}{\sigma({\mathscr B})}
\,=\frac{\,\pm\;{\bf D}({\bf b})\,{\bf L}({\bf b},\,\lambda)\,-\,0\,}{\sigma({\mathscr B})}
\,=\,\left\{\frac{\,\pm\,{\bf D}({\bf b})\,}{\sigma({\mathscr B})}\right\}\,{\bf L}({\bf b},\,\lambda)
\,=\,{\bf L}({\bf b},\,\lambda)\,.\label{standi}
\end{align}
This confirms the standard scores derived in the equations (\ref{nonvar-a}) and (\ref{nonvar-b}) of the previous subsection.

\eject

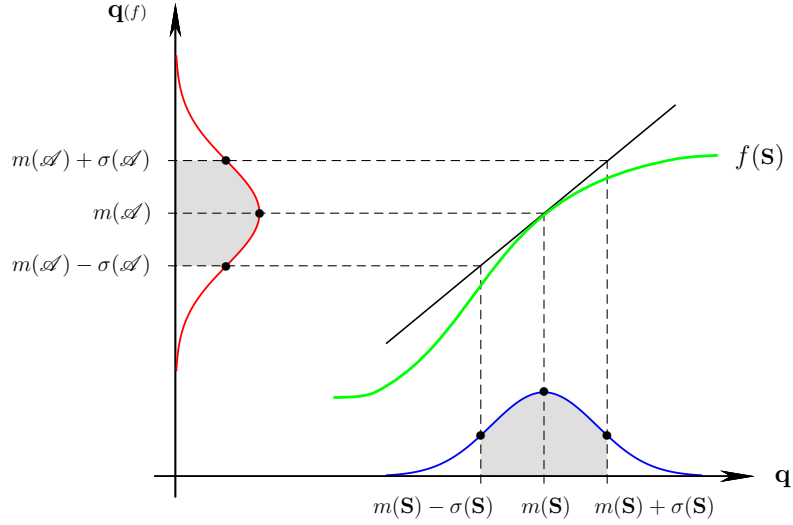
\begin{figure}
\hrule
\scalebox{0.7}{
\psset{yunit=2cm,xunit=2}
\begin{pspicture}(-0.6,-0.5)(5,5)

\uput[0](-0.73,4.4){{\Large ${\bf q}$}${(f)}$}

\uput[0](-1.64,3.0){\large ${m({\mathscr A})+\sigma({\mathscr A})}$}

\uput[0](-0.87,2.5){\large ${m({\mathscr A})}$}

\uput[0](-1.64,2.0){\large ${m({\mathscr A})-\sigma({\mathscr A})}$}

\uput[-90](5.765,0.15){\Large ${\bf q}$}

\uput[-90](5.55,3.25){\Large ${f}$({\large{\bf S}})}

\uput[-90](2.45,-0.1){\large ${m({\bf S})-\sigma({\bf S})}$}

\uput[-90](3.52,-0.1){\large ${m({\bf S})}$}

\uput[-90](4.55,-0.1){\large ${m({\bf S})+\sigma({\bf S})}$}

\pscustom[linestyle=none,fillstyle=solid,fillcolor=gray!25]{%
\psGauss[sigma=0.5,mue=3.5,linewidth=1pt]{2.9}{4.1}
      \psline(4.1,0.0)(2.9,0.0)}

\psGauss[sigma=0.5,mue=3.5,linewidth=1pt,linecolor=blue]{2.0}{5.0}

\uput[-90](2.9,0.519){\large ${\bullet}$}

\uput[-90](3.5,0.931){\large ${\bullet}$}

\uput[-90](4.1,0.519){\large ${\bullet}$}

\psline[linewidth=0.2mm,linestyle=dashed]{-}(2.9,-0.07)(2.9,2.0)

\psline[linewidth=0.2mm,linestyle=dashed]{-}(4.1,-0.07)(4.1,3.0)

\begin{rotate}{-90}
\pscustom[linestyle=none,fillstyle=solid,fillcolor=gray!25]{%
\psGauss[sigma=0.5,mue=-2.5,linewidth=1pt,fillstyle=solid,fillcolor=gray!25]{-3.00}{-2.0}%
      \psline(-2.0,0.0)(-3.0,0.0)}
\end{rotate}

\psaxes[labels=none,ticksize=0pt,arrowinset=0.3,arrowsize=3pt 4,arrowlength=3]{->}(0,0)(-0.2,-0.2)(5.5,4.5)

\psline[linewidth=0.2mm,linestyle=dashed]{-}(-0.07,2.0)(2.9,2.0)

\psline[linewidth=0.2mm,linestyle=dashed]{-}(-0.07,3.0)(4.1,3.0)

\begin{rotate}{-90}
\psGauss[sigma=0.5,mue=-2.5,linewidth=1pt,linecolor=red]{-4.0}{-1.0}%
\end{rotate}

\uput[-90](0.48,3.134){\large ${\bullet}$}

\uput[-90](0.8,2.625){\large ${\bullet}$}

\uput[-90](0.48,2.125){\large ${\bullet}$}

\psline[linewidth=0.3mm]{-}(2.0,1.26)(4.75,3.53)

\psline[linewidth=0.2mm,linestyle=dashed]{-}(3.5,-0.07)(3.5,2.5)

\psline[linewidth=0.2mm,linestyle=dashed]{-}(-0.07,2.5)(3.5,2.5)

\pscurve[linewidth=1.5pt,linecolor=green]{-}(1.5,0.75)(1.9,0.8)(3.5,2.483)(4.0,2.79)(4.75,3.015)(5.15,3.05)

\end{pspicture}}
\vspace{0.25cm}
\hrule
\caption{Propagation of the 68\% probability interval from a random bivector ${\bf S}$ to a scalar ${\mathscr A}$
within the parallelized 3-sphere.}
\label{fig-88}
\vspace{0.3cm}
\hrule
\end{figure}

So far I have assumed that randomness in the measurement results ${{\mathscr A}{\bf (a, \lambda)}}$ and
${{\mathscr B}{\bf (b, \lambda)}}$ originates entirely from the initial state ${\lambda}$ representing the orientation of
the 3-sphere. In other words, I have assumed that the local interaction of the detector ${\bf D(a)}$ with the random
spin ${\bf L(a, \lambda)}$ does not introduce additional randomness in the measurement result ${{\mathscr A}{\bf (a, \lambda)}}$.
Any realistic interaction between ${\bf D(a)}$ and ${\bf L(a, \lambda)}$, however, would inevitably introduce such a randomness,
of purely local, experimental origin. We can model this randomness
by introducing an additional random variable, say $r_{\bf a}\in[\,0,\,1]$, not dependent on ${\lambda}$. Physically we
can think of $r_{\bf a}$ as an alignment parameter between the detector bivector ${\bf D(a)}$ and the spin bivector
${\bf L(a, \lambda)}$, with $r_{\bf a}=1$ representing the perfect alignment. Clearly, introduction of this additional
random parameter will make all the bivectors and quaternions unnormalized, and the corresponding probability density function
(\ref{probdensi-8}) would then represent a Gaussian distribution---provided we also assume that the orientation
${\lambda=\pm\,1}$ of $S^3$ itself is distributed non-uniformly between its values $+1$ and $-1$.
Moreover, although the measurement
results would then fall within the range ${-\,1\,\leq\,{\mathscr A}({\bf a},\,\lambda)\,\leq\,+\,1}$, their mean value would
be zero for a uniformly distributed ${\lambda}$, since the mean value of the product of the independent random variables
${r_{\bf a}}$ and ${\lambda}$ would then be the product of their mean values:
\begin{equation}
m(r_{\bf a}\,\lambda)\;=\;m(r_{\bf a})\,m(\lambda).
\end{equation}
More importantly, the standard scores computed above in equations (\ref{standj}) and (\ref{standi}) would not be affected by this
more realistic random process ${r_{\bf a}\,\lambda}$---at least for the special case of uniformly distributed ${\lambda}$---because
they involve the ratios of the corresponding raw scores and standard deviations centered about the mean values
${m(r_{\bf a}\,\lambda)=0=m(r_{\bf b}\,\lambda)}$:
\begin{align}
{A}({\bf a},\,\lambda)
\,&=\frac{\,\pm\,{\mathscr A}({\bf a},\,{\lambda})\,-\,
{\overline{{\mathscr A}({\bf a},\,{\lambda})}}}{\sigma({\mathscr A})}
\,=\frac{\,\pm\;r_{\bf a}\,{\bf D}({\bf a})\,{\bf L}({\bf a},\,\lambda)\,-\,0\,}{\sigma({\mathscr A})}
\,=\,\left\{\frac{\,\pm\,r_{\bf a}\,{\bf D}({\bf a})\,}{\sigma({\mathscr A})}\right\}\,{\bf L}({\bf a},\,\lambda)
\,=\,{\bf L}({\bf a},\,\lambda) \label{standk}\\
\text{and}\;\;\;{B}({\bf b},\,\lambda)
\,&=\frac{\,\pm\,{\mathscr B}({\bf b},\,{\lambda})\,-\,
{\overline{{\mathscr B}({\bf b},\,{\lambda})}}}{\sigma({\mathscr B})}
\,=\frac{\,\pm\;r_{\bf b}\,{\bf D}({\bf b})\,{\bf L}({\bf b},\,\lambda)\,-\,0\,}{\sigma({\mathscr B})}
\,=\,\left\{\frac{\,\pm\,r_{\bf b}\,{\bf D}({\bf b})\,}{\sigma({\mathscr B})}\right\}\,{\bf L}({\bf b},\,\lambda)
\,=\,{\bf L}({\bf b},\,\lambda)\,.\label{standl}
\end{align}

Let us now try to understand the propagation of error within ${S^3}$ from this physically more realistic perspective.
To this end, let the random variable ${{\bf q}(\psi,\,{\bf a},\,\lambda)\in S^3(\lambda)}$ be such that the
measurement results ${{\mathscr A}({\bf a},\,\lambda)\in[-1,\,+1]}$ remain as before, but the bivectors
${{\bf L}({\bf a},\,\lambda)}$ are subject to a random process ${r_{\bf a}\,\lambda}$ such that
${{\bf S}({\bf a},\,\lambda,\,r_{\bf a})=r_{\bf a}\,{\bf L}({\bf a},\,\lambda)}$ with ${r_{\bf a}\in[{\hspace{1pt}}0,\,1]}$.
Then the mean value ${m({\bf S})}$ and standard deviation ${\sigma({\bf S})}$ of ${\bf S}$ would be a bivector and a scalar:
\begin{align}
m({\bf S}) & \,=\;\text{a bivector}\;\;\;\;\;\;\;\; \notag \\
\text{and}\;\;\;\sigma({\bf S})& \,=\;\text{a scalar}.
\end{align}
If we now take the detector bivector to be ${{\bf D}({\bf a})=I\cdot{\bf a}}$ as before, then the
measurement results can be identified as ${\,-1 \leq {\mathscr A} = {\bf D}\,{\bf S} \leq +1\,}$ so
that ${m({\mathscr A})\geq 0}$. Since ${\bf D}$ is a non-random bivector,
errors generated within ${\mathscr A}$ by the random process ${r_{\bf a}\,\lambda}$ would stem
entirely from the random bivector ${\bf S}$, and propagate linearly. In other words,
the standard deviations within the random number ${\mathscr A}$ due to the random
process ${r_{\bf a}\,\lambda}$ would be given by
\begin{equation}
\sigma({\mathscr A})={\bf D}\,\sigma({\bf S}).
\end{equation}
But since ${\sigma({\bf S})}$, as we noted, is a scalar, the typical error 
${\sigma({\mathscr A})}$ generated within ${\mathscr A}$ due to the
random process ${r_{\bf a}\,\lambda}$ is a bivector. The standardized variable
(which must be used to compare the raw scores ${\mathscr A}$ with other raw
scores ${{\mathscr B}\,}$) is thus also a bivector: ${A:={\mathscr A}/{\sigma({\mathscr A})}=\text{scalar}\times{\bf S}}$.

As straightforward as it is, the above conclusion may seem unusual.
It is important to recall, however, that in geometric algebra both scalars and bivectors are treated
on equal footing \ocite{Clifford}\ocite{Hestenes}. They both behave as real-valued c-numbers, albeit of different grades.
To appreciate the consistency and naturalness of the above conclusion, let
\begin{equation}
{\mathscr A}\,=\,f({\bf S})\,=\,{\bf D}\,{\bf S}\label{defofa-0}
\end{equation}
be a continuous function generated by the geometric product of two bivectors ${{\bf D}({\bf a})}$
and ${{\bf S}({\bf a},\,\lambda,\,r_{\bf a})}$ as before.
The natural question then is: How does a typical error in ${\bf S}$ governed by the probability density
(\ref{probdensi-8})---which can be represented by the 68\% probability interval
\begin{equation}
\left[\,m({\bf S})-\sigma({\bf S}),\;m({\bf S})+\sigma({\bf S})\,\right]\label{int-1no}
\end{equation}
as shown in the Fig.${\,}$\ref{fig-88}---propagate from the random bivector ${\bf S}$ to the random
scalar ${\mathscr A}$, through the function ${f({\bf S}) = {\bf D}\,{\bf S}}$? To answer this question
we note that the two end points of the interval (\ref{int-1no}) represent two points, say
${{\bf q}^-}$ and ${{\bf q}^+}$, of the 3-sphere, which is a Riemannian manifold. The
geometro-algebraic distance between the points ${{\bf q}^-}$ and ${{\bf q}^+}$ can therefore be defined, say, as
\begin{equation}
d{\hspace{-1pt}}\left({\bf q}^-\!,\,{\bf q}^+\right)
\,=\,\left({\bf q}^- -\,{\bf q}^+\right)\times \text{sign}\!\left({\bf q}^- -\,{\bf q}^+\right).
\end{equation}
Moreover, from definition (\ref{defofa-0}) of ${\mathscr A}$ and a first-order Taylor expansion of
the function ${f({\bf S})}$ about the point ${{\bf S} = m({\bf S})}$ we obtain
\begin{equation}
{\mathscr A}\,=\,f(m({\bf S}))\,+\,\frac{\partial f}{\partial {\bf S}}\bigg|_{{\bf S}\,=\;m({\bf S})}
({\bf S}\,-\,m({\bf S}))\,+\,\dots \label{Tylore}
\end{equation}
Now it is evident that the slope ${{\partial f}/{\partial {\bf S}}={\bf D}}$ of this line is a constant.
Therefore the mean ${m({\mathscr A})}$ and the standard deviation ${\sigma({\mathscr A})}$ of the distribution of
${{\mathscr A}\!}$ can be obtained by setting ${{\bf S}=m({\bf S})}$ and ${{\bf S}=\sigma({\bf S})}$:
\begin{align}
m({\mathscr A})\,&=\,f(m({\bf S}))\,=\,{\bf D}\, m({\bf S})\,=\,\text{a scalar} \\
\text{and}\;\;\sigma({\mathscr A})\,&=\,\frac{\partial f}{\partial {\bf S}}\,\sigma({\bf S})
\,=\,{\bf D}\,\sigma({\bf S})\,=\,\text{a bivector}.
\end{align}
The probability distribution of ${\mathscr A}$ is thus represented by the 68\% interval
\begin{equation}
\left[\,m({\mathscr A})-\sigma({\mathscr A}),\;m({\mathscr A})+\sigma({\mathscr A})\,\right].\label{int-2no}
\end{equation}
If we now set ${r_{\bf a}=1}$ and thereby assume that ${\bf S}$ is in fact
the unit bivector ${\bf L}$ with a vanishing mean, then we have ${m({\mathscr A})=0}$
and ${\sigma({\mathscr A})=\pm\,{\bf D}}$, as in equation (\ref{a30}) above.

\makeatletter
\renewcommand*{\rightarrowfill@}{%
   \arrowfill@\relbar\relbar\chemarrow}
\makeatother

Finally, it is instructive to note that, geometrically, the propagation of error within ${S^3}$
is equivalent to a simple change in the perspective (cf. Fig.${\,}$\ref{fig-88}):  
\begin{equation}
S^3\ni\,
  \underbrace{
    \overbrace{m({\bf S})}^\text{bivector} \;\pm\;\,
    \overbrace{\sigma({\bf S})}^\text{scalar}
   }_\text{quaternion}
\;\;\xrightarrow[]{\text{\;\;\;\;\;\;${f({\bf S})}$\;\;\;\;\;}}\;\;
\underbrace{
    \overbrace{m({\mathscr A})}^\text{scalar} \,\;\pm\;
    \overbrace{\sigma({\mathscr A})}^\text{bivector}
   }_\text{quaternion}.
\end{equation}
In particular, the probability density of the scalar ${\mathscr A}$ over ${S^3}$ corresponding to interval (\ref{int-2no}) is
equivalent to that of the bivector ${\bf S}$ over ${S^3}$ corresponding to interval (\ref{int-1no}).
With this, we are now ready to derive the EPR-Bohm correlations.

\subsection{Derivation of Pair Correlations Among the Points of ${S^3}$}

We begin by noting that, according to my model, EPR-Bohm correlations are correlations among the points of a parallelized
3-sphere \ocite{Christian}. Now, since we have assumed that initially there was 50/50 chance between the right-handed and
left-handed orientations of the 3-sphere ({\it i.e.}, equal chance between the initial states ${{\lambda}=+\,1}$ and
${{\lambda}=-\,1}$), the expectation values of the raw scores ${{\mathscr A}({\bf a},\,{\lambda})}$ and
${{\mathscr B}({\bf b},\,{\lambda})}$ vanish identically. On the other hand, as discussed above, the correlation between
these raw scores (or their first product moment coefficient {\it \`a la} Pearson \ocite{scores-1}) can
be obtained only by computing the covariance between the corresponding standardized variables
${{A}({\bf a},\,{\lambda})}$ and ${{B}({\bf b},\,{\lambda})}$, which gives
\begin{align}
{\cal E}({\bf a},\,{\bf b})\,
&=\,\lim_{\,n\,\gg\,1}\left[\frac{1}{n}\sum_{k\,=\,1}^{n}\,A({\bf a},\,{\lambda}^k)\,B({\bf b},\,{\lambda}^k)\right] \notag \\
&=\lim_{\,n\,\gg\,1}\left[\frac{1}{n}\sum_{k\,=\,1}^{n}\,
\left\{\,a_{\mu}\;L_{\mu}(\lambda^k)\,\right\}\,\left\{\,b_{\nu}\;L_{\nu}(\lambda^k)\,\right\}\right] \notag \\
&=\,-\,g_{\mu\nu}\,a_{\mu}\,b_{\nu}\,-\lim_{\,n\,\gg\,1}\left[\frac{1}{n}\sum_{k\,=\,1}^{n}\,
\left\{\,\epsilon_{\mu\nu\rho}\;a_{\mu}\,b_{\nu}\;L_{\rho}(\lambda^k)\,\right\}\right] \notag \\
&=\,-\,g_{\mu\nu}\,a_{\mu}\,b_{\nu}\,-\lim_{\,n\,\gg\,1}\left[\frac{1}{n}\sum_{k\,=\,1}^{n}\,
\lambda^k\right]\left\{\,\epsilon_{\mu\nu\rho}\;a_{\mu}\,b_{\nu}\;D_{\rho}\,\right\} \notag \\
&=\,-\,g_{\mu\nu}\,a_{\mu}\,b_{\nu}\,-\,0\,, \label{corwedid}
\end{align}
where I have used algebra defined in (\ref{wh-o88}) and the relation (\ref{OJS}).
Consequently, as explained in the paragraph just below Eq.${\,}$(\ref{stan}),
when the raw scores ${{\mathscr A}=\pm\,1}$ and ${{\mathscr B}=\pm\,1}$ are
compared, their product moment will inevitably yield
\begin{equation}
{\cal E}({\bf a},\,{\bf b})\,=\lim_{\,n\,\gg\,1}\left[\frac{1}{n}\sum_{k\,=\,1}^{n}\,
{\mathscr A}({\bf a},\,{\lambda}^k)\;{\mathscr B}({\bf b},\,{\lambda}^k)\right]\,
=\,-\,g_{\mu\nu}\,a_{\mu}\,b_{\nu}\,, \label{darcor}
\end{equation}
since the correlation between the raw scores ${{\mathscr A}}$ and ${\mathscr B}$ is
equal to covariance between the standard scores ${A}$ and${\;B}$.

So far in this section we have put no restrictions on the metric tensor, which, in the normal coordinates centered at a point
of ${S^3}$ would be of the form
\begin{equation}
g_{\mu\nu}(x)\,=\,\delta_{\mu\nu}\,-\,\frac{1}{3}\;
{\mathscr R}_{\,\alpha\,\mu\,\nu\,\gamma}\,x^{\alpha}\,x^{\gamma}\,+\,O\left(|x|^3\right).
\end{equation}
In other words, the algebra (\ref{wh-o88}) we have used in the derivation of correlation (\ref{darcor}) is a general
Clifford algebra, with no restrictions placed on the quadratic form \ocite{Frankel}.
On the other hand, if the codomain of the measurement functions ${{\mathscr A}({\bf a},\,{\lambda})}$ is taken to be
a parallelized 3-sphere, then the above metric tensor specializes to the Euclidean metric ${\delta_{\mu\nu}}$, because the
Riemann curvature tensor of a parallelized 3-sphere vanishes, inducing a non-vanishing torsion \ocite{Chris-macro}. This case
corresponds to the geometry of the group SU(2) and specializes the correlation (\ref{darcor}) to exhibit maximum strength:
\begin{equation}
{\cal E}({\bf a},\,{\bf b})\,=\,-\,g_{\mu\nu}\,a_{\mu}\,b_{\nu}
\,\longrightarrow\,-\,\delta_{\mu\nu}\,a_{\mu}\,b_{\nu}\,=\,-\,\cos\eta_{{\bf a}{\bf b}}\,, \label{robisanan}
\end{equation}
which in turn manifests the sensitivity of ${{\mathscr A}({\bf a},\,{\lambda})}$ and
${{\mathscr B}({\bf b},\,{\lambda})}$ to spinorial sign changes.
To appreciate the significance of these changes \ocite{Chris-macro},
recall from subsection II${\,}$A that a parallelized 3-sphere is a set of unit quaternions of the form
\begin{equation}
{\bf q}(\psi,\,{\bf r})\,=\,\cos\frac{\psi}{2}\,+\,{\boldsymbol\beta}({\bf r})\,\sin\frac{\psi}{2}\,, \label{defi-2-b}
\end{equation}
with ${\psi}$ being the rotation angle. It is easy to check that ${{\bf q}(\psi,\,{\bf r})}$ respects the following rotational
symmetries:
\begin{align}
{\bf q}(\psi+2\kappa\pi,\,{\bf r})\,&=\,-\,{\bf q}(\psi,\,{\bf r})\,\;\;\text{for}\;\,\kappa=1,3,5,7,\dots \label{clacct}\\
{\bf q}(\psi+4\kappa\pi,\,{\bf r})\,&=\,+\,{\bf q}(\psi,\,{\bf r})\,\;\;\text{for}\;\,\kappa=0,1,2,3,\dots \label{non-clacct}
\end{align}
Thus ${{\bf q}(\psi,\,{\bf r})}$ correctly represents the state of a body that returns to itself only after even multiples
of a ${2\pi}$ rotation.

It is very important to appreciate that the strong correlation derived in (\ref{robisanan}) are correlation among the points of a
parallelized 3-sphere, ${S^3}$, taken as the codomain of the measurement functions ${{\mathscr A}({\bf r},\,\lambda)}$. Thus the
strength and the very existence of the EPR-Bohm correlation (or of {\it any} correlation for that matter) stem entirely from the
topological properties of the codomain of the measurement functions ${{\mathscr A}({\bf r},\,\lambda)}$. Had we chosen the
codomain of ${{\mathscr A}({\bf r},\,\lambda)}$ to be any manifold other than a parallelized 3-sphere, the resulting correlation
would not have been as strong as ${\,-\cos\eta_{{\bf a}{\bf b}}}$.

\subsection{Derivation of Upper Bound Exceeding the Bell-CHSH Bound}

Returning to the expectation value (\ref{darcor}) in its most general form we can now proceed to derive the Bell-CHSH-type
bound on possible correlations \ocite{Christian}\ocite{bounds}. To this end, consider four observation axes,
${\bf a}$, ${\bf a'}$, ${\bf b}$, and ${\bf b'}$, for the standard EPR-Bohm experiment.
Then the corresponding CHSH string of expectation values \ocite{Christian}, namely the coefficient
\begin{equation}
{\cal S}({\bf a},\,{\bf a'},\,{\bf b},\,{\bf b'})\,:=\,{\cal E}({\bf a},\,{\bf b})\,+\,{\cal E}({\bf a},\,{\bf b'})\,+\,
{\cal E}({\bf a'},\,{\bf b})\,-\,{\cal E}({\bf a'},\,{\bf b'})\,, \label{B1-11}
\end{equation}
would be bounded by the constant ${2\sqrt{2}}$, as discovered by Tsirel'son within the setting of Clifford algebra applied
to quantum mechanics in general \ocite{Christian}\ocite{bounds}. Here each of the joint expectation values of the raw scores
${{\mathscr A}({\bf a},\,{\lambda})=\pm\,1}$ and ${{\mathscr B}({\bf b},\,{\lambda})=\pm\,1}$ are defined as
\begin{equation}
{\cal E}({\bf a},\,{\bf b})\,=\lim_{\,n\,\gg\,1}\left[\frac{1}{n}\sum_{k\,=\,1}^{n}\,
{\mathscr A}({\bf a},\,{\lambda}^k)\;{\mathscr B}({\bf b},\,{\lambda}^k)\right],\label{exppeu}
\end{equation}
with the binary numbers such as ${{\mathscr A}({\bf a},\,{\lambda})}$ defined by the limit
\begin{equation}
S^3\ni\pm\,1\,=\,{\mathscr A}({\bf a},\,\lambda)\,
=\,\lim_{{\psi}\rightarrow\,{2\kappa\pi}}\,\left\{\,{\bf q}(\psi,\,{\bf a},\,\lambda)\,\right\}\,
=\,-\,{\bf D}({\bf a})\,{\bf L}({\bf a},\,\lambda).
\end{equation}
Thus ${{\mathscr A}({\bf a},\,{\lambda})}$ and  ${{\mathscr B}({\bf b},\,{\lambda})}$ are points of a parallelized 3-sphere
and ${{\cal E}({\bf a},\,{\bf b})}$ evaluated in (\ref{exppeu}) gives correlation between such points of the 3-sphere
\ocite{Christian}. The correct value of the correlation, however, cannot be obtained without appreciating the fact that the
number ${{\mathscr A}({\bf a},\,{\lambda})=\pm\,1}$ is defined as a product of a ${\lambda}$-{\it in}-dependent constant,
namely ${-\,{\bf D}({\bf a})}$, and a ${\lambda}$-dependent random
variable, namely ${{\bf L}({\bf a},\,\lambda)}$. Thus the correct value of the correlation
is obtained by calculating the covariance of the corresponding standardized variables
\begin{align}
A_{\bf a}({\lambda})\,\equiv\,A({\bf a},\,\lambda)\,&=\,{\bf L}({\bf a},\,\lambda) \label{dumtit-1} \\
\text{and}\;\;\;B_{\bf b}({\lambda})\,\equiv\,B({\bf b},\,\lambda)\,&=\,{\bf L}({\bf b},\,\lambda)\,, \label{dumtit-2}
\end{align}
as we discussed just below equation (\ref{stan}). In other words, correlation between the raw scores
${{\mathscr A}({\bf a},\,{\lambda})}$ and ${{\mathscr B}({\bf b},\,{\lambda})}$ is obtained by calculated
the covariance between the standard scores ${{A}({\bf a},\,{\lambda})}$ and ${{B}({\bf b},\,{\lambda})}$,
as in equation (\ref{corwedid}) above:
\begin{equation}
{\cal E}({\bf a},\,{\bf b})\,=\lim_{\,n\,\gg\,1}\left[\frac{1}{n}\sum_{k\,=\,1}^{n}\,
{A}({\bf a},\,{\lambda}^k)\;{B}({\bf b},\,{\lambda}^k)\right]\,=\,-\,g_{\mu\nu}\,a_{\mu}\,b_{\nu}\,.\label{stand-exppeu}
\end{equation}
The correlation between the raw scores is thus necessarily equal to
the covariance between the standard scores:
\begin{equation}
\boxed{\;{\cal E}({\bf a},\,{\bf b})\,=\lim_{\,n\,\gg\,1}\left[\frac{1}{n}\sum_{k\,=\,1}^{n}\,
{\mathscr A}({\bf a},\,{\lambda}^k)\;{\mathscr B}({\bf b},\,{\lambda}^k)\right]
\,=\lim_{\,n\,\gg\,1}\left[\frac{1}{n}\sum_{k\,=\,1}^{n}\,
{A}({\bf a},\,{\lambda}^k)\;{B}({\bf b},\,{\lambda}^k)\right]\,=\,-\,g_{\mu\nu}\,a_{\mu}\,b_{\nu}\,.\,} \label{stoppsa}
\end{equation}
Using this identity we can now rewrite the CHSH string of expectation values (\ref{B1-11}) in two equivalent expressions,
\begin{align}
{\cal S}({\bf a},\,{\bf a'},\,{\bf b},\,{\bf b'})\,=\,\lim_{\,n\,\gg\,1}\Bigg[\frac{1}{n}\sum_{k\,=\,1}^{n}\,
&{\mathscr A}_{\bf a}({\lambda}^k)\,{\mathscr B}_{\bf b}({\lambda}^k)\Bigg]\,+\,
\lim_{\,n\,\gg\,1}\Bigg[\frac{1}{n}\sum_{k\,=\,1}^{n}\,
{\mathscr A}_{\bf a}({\lambda}^k)\,{\mathscr B}_{\bf b'}({\lambda}^k)\Bigg]\, \notag \\
&+\,\lim_{\,n\,\gg\,1}\Bigg[\frac{1}{n}\sum_{k\,=\,1}^{n}\,
{\mathscr A}_{\bf a'}({\lambda}^k)\,{\mathscr B}_{\bf b}({\lambda}^k)\Bigg]\,-\,
\lim_{\,n\,\gg\,1}\Bigg[\frac{1}{n}\sum_{k\,=\,1}^{n}\,
{\mathscr A}_{\bf a'}({\lambda}^k)\,{\mathscr B}_{\bf b'}({\lambda}^k)\Bigg] \label{probnonint}
\end{align}
and
\begin{align}
\;\;\;\;\;\;\;{\cal S}({\bf a},\,{\bf a'},\,{\bf b},\,{\bf b'})\,=\,\lim_{\,n\,\gg\,1}\Bigg[\frac{1}{n}\sum_{k\,=\,1}^{n}\,
&{A}_{\bf a}({\lambda}^k)\,{B}_{\bf b}({\lambda}^k)\Bigg]\,+\,
\lim_{\,n\,\gg\,1}\Bigg[\frac{1}{n}\sum_{k\,=\,1}^{n}\,
{A}_{\bf a}({\lambda}^k)\,{B}_{\bf b'}({\lambda}^k)\Bigg]\, \notag \\
&+\,\lim_{\,n\,\gg\,1}\Bigg[\frac{1}{n}\sum_{k\,=\,1}^{n}\,
{A}_{\bf a'}({\lambda}^k)\,{B}_{\bf b}({\lambda}^k)\big\}\Bigg]\,-\,
\lim_{\,n\,\gg\,1}\Bigg[\frac{1}{n}\sum_{k\,=\,1}^{n}\,
{A}_{\bf a'}({\lambda}^k)\,{B}_{\bf b'}({\lambda}^k)\Bigg]. \label{probnonrrint}
\end{align}
Our goal now is to find the upper bound on these strings of expectation values. To this end, we first note that the four pairs of
measurement results occurring in the above expressions do not all occur at the same time. Let us, however, conform to the usual
assumption of counterfactual definiteness and pretend that they do occur at the same time, at least counterfactually, with equal
distribution. This assumption allows us to simplify the above expressions as
\begin{equation}
{\cal S}({\bf a},\,{\bf a'},\,{\bf b},\,{\bf b'})\,=\,\lim_{\,n\,\gg\,1}\Bigg[\frac{1}{n}\sum_{k\,=\,1}^{n}\,\big\{
{\mathscr A}_{\bf a}({\lambda}^k)\,{\mathscr B}_{\bf b}({\lambda}^k)\,+\,
{\mathscr A}_{\bf a}({\lambda}^k)\,{\mathscr B}_{\bf b'}({\lambda}^k)\,+\,
{\mathscr A}_{\bf a'}({\lambda}^k)\,{\mathscr B}_{\bf b}({\lambda}^k)\,-\,
{\mathscr A}_{\bf a'}({\lambda}^k)\,{\mathscr B}_{\bf b'}({\lambda}^k)\big\}\Bigg] \label{probnontint}
\end{equation}
and
\begin{equation}
{\cal S}({\bf a},\,{\bf a'},\,{\bf b},\,{\bf b'})\,=\,\lim_{\,n\,\gg\,1}\Bigg[\frac{1}{n}\sum_{k\,=\,1}^{n}\,\big\{
A_{\bf a}({\lambda}^k)\,B_{\bf b}({\lambda}^k)\,+\,
A_{\bf a}({\lambda}^k)\,B_{\bf b'}({\lambda}^k)\,+\, A_{\bf a'}({\lambda}^k)\,B_{\bf b}({\lambda}^k)\,-\,
A_{\bf a'}({\lambda}^k)\,B_{\bf b'}({\lambda}^k)\big\}\Bigg]. \label{probnonint-b}
\end{equation}

The obvious question now is: Which of these two expressions should we evaluate to obtain the correct bound on 
${{\cal S}({\bf a},\,{\bf a'},\,{\bf b},\,{\bf b'})}$? Clearly, in view of the identity (\ref{stoppsa}) both expressions
would give one and the same answer \ocite{Christian}. Thus it should not matter which of the two expressions we use to evaluate
the bound. But it is also clear from the discussion in subsections II${\,}$C and II${\,}$D that the correct bound on the expression
(\ref{probnontint}) involving the raw scores ${\mathscr A}$ and ${\mathscr B}$ can only be obtained by evaluating the expression
(\ref{probnonint-b}) involving the standard scores ${A}$ and ${B}$. Stated differently, if we tried to obtain the bound on
${{\cal S}({\bf a},\,{\bf a'},\,{\bf b},\,{\bf b'})}$ by disregarding how the measurement results have been
generated in the model statistically, then we would end up getting a wrong answer. By following the Bell-CHSH reasoning blindly
Weatherall ends up making such a mistake. In the end ${{\cal S}({\bf a},\,{\bf a'},\,{\bf b},\,{\bf b'})}$
is a functional of a {\it random} variable, and as such proper statistical procedure tailored to my model must be employed for
its correct evaluation. This is an important point and the reader is urged to review the discussions in
subsections II${\,}$C and II${\,}$D once again to appreciate its full significance.

With these remarks in mind we proceed to obtain the upper bound on ${{\cal S}({\bf a},\,{\bf a'},\,{\bf b},\,{\bf b'})}$
by evaluating the expression (\ref{probnonint-b}) as follows. Since the standard scores
${A_{\bf a}({\lambda})={\bf L}({\bf a},\,\lambda)}$ and ${B_{\bf b}({\lambda})={\bf L}({\bf b},\,\lambda)}$ appearing in
this expression represent two independent equatorial points of the 3-sphere, we can take them to belong
to two disconnected ``sections'' of ${S^3}$ ({\it i.e.}, two disconnected 2-spheres within ${S^3}$), satisfying
\begin{equation}
\left[\,A_{\bf r}({\lambda}),\,B_{\bf r'}({\lambda})\,\right]\,=\,0\,
\;\;\;\forall\;\,{\bf r}\;\,{\rm and}\;\,{\bf r'}\,\in\,{\rm I\!R}^3,\label{com}
\end{equation}
which is equivalent to anticipating a null outcome along the direction ${{\bf r}\times{\bf r'}}$ exclusive
to both ${\bf r}$ and ${\bf r'}$.
If we now square the integrand of equation (\ref{probnonint-b}), use the above commutation relations, and use the fact
that all bivectors square to ${-1}$, then the absolute value of ${{\cal S}({\bf a},\,{\bf a'},\,{\bf b},\,{\bf b'})}$ leads
to the following variance inequality \ocite{Christian}:
\begin{align}
|{\cal S}({\bf a},\,{\bf a'},\,{\bf b},\,{\bf b'})|\,&=\,
|{\cal E}({\bf a},\,{\bf b})\,+\,{\cal E}({\bf a},\,{\bf b'})\,+\,
{\cal E}({\bf a'},\,{\bf b})\,-\,{\cal E}({\bf a'},\,{\bf b'})|\, \notag \\
&\leqslant\sqrt{\lim_{\,n\,\gg\,1}\left[\frac{1}{n}\sum_{k\,=\,1}^{n}\,
\big\{\,4\,+\,4\,{\cal T}_{\,{\bf a\,a'}}({\lambda}^k)\,{\cal T}_{\,{\bf b'\,b}}({\lambda}^k)\,\big\}\right]},\label{yever}
\end{align}
where the classical commutators
\begin{equation}
{\cal T}_{\,{\bf a\,a'}}(\lambda):=\frac{1}{2}\left[\,A_{\bf a}(\lambda),\,A_{\bf a'}(\lambda)\right]
\,=\,-\,A_{{\bf a}\times{\bf a'}}(\lambda) \label{aa-potorsion-666}
\end{equation}
and
\begin{equation}
{\cal T}_{\,{\bf b'\,b}}(\lambda)
:=\frac{1}{2}\left[\,B_{\bf b'}(\lambda),\,B_{\bf b}(\lambda)\right]\,=\,-\,B_{{\bf b'}\times{\bf b}}(\lambda)\label{bb-potor}
\end{equation}
are the geometric measures of the torsion within ${S^3}$. Thus, it is the non-vanishing torsion ${\cal T}$ within the parallelized
3-sphere---the parallelizing torsion which makes its Riemann curvature tensor vanish---that is ultimately responsible for
the strong quantum correlation \ocite{Christian}\ocite{what}. We can see this at once from Eq.${\,}$(\ref{yever}) by setting
${{\cal T}=0}$, and in more detail as follows: Using definitions (\ref{dumtit-1}) and (\ref{dumtit-2}) for ${A_{\bf a}({\lambda})}$
and ${B_{\bf b}({\lambda})}$ and making a repeated use of the bivector identity
\begin{equation}
{\bf L}({\bf a},\,\lambda)\,{\bf L}({\bf a'},\,\lambda)\,=\,-\,{\bf a}\cdot{\bf a'}\,-\,
{\bf L}({\bf a}\times{\bf a'},\,\lambda)
\end{equation}
specialized for the metric ${g_{{\mu}{\nu}}=\delta_{{\mu}{\nu}}}$ on ${S^3}$, the above inequality
for ${{\cal S}({\bf a},\,{\bf a'},\,{\bf b},\,{\bf b'})}$ can be further simplified to
\begin{align}
|{\cal S}({\bf a},\,{\bf a'},\,{\bf b},\,{\bf b'})|\,
&\leqslant\sqrt{\!4-4\,({{\bf a}}\times{{\bf a}'})\cdot({{\bf b}'}\times{{\bf b}})-
4\!\lim_{\,n\,\gg\,1}\left[\frac{1}{n}\sum_{k\,=\,1}^{n}{{\bf L}}({\bf z},\,\lambda^k)\right]} \notag \\
&\leqslant\sqrt{\!4-4\,({{\bf a}}\times{\bf a'})\cdot({\bf b'}\times{{\bf b}})-
4\!\lim_{\,n\,\gg\,1}\left[\frac{1}{n}\sum_{k\,=\,1}^{n}\lambda^k\right]{{\bf D}}({\bf z})} \notag \\
&\leqslant\,2\,\sqrt{\,1-({{\bf a}}\times{\bf a'})
\cdot({\bf b'}\times{{\bf b}})\,-\,0\,}\,,\label{before-opppo-666}
\end{align}
where ${{\bf z}=({\bf a}\times{\bf a'})\times({\bf b'}\times{\bf b})}$, and---as before---I have used the relation (\ref{OJS})
between ${{\bf L}({\bf z},\,\lambda)}$ and ${{\bf D}({\bf z})}$ from subsection II${\,}$B.
Finally, by noticing that ${({\bf a}\times{\bf a'})\cdot({\bf b'}\times{\bf b})}$ is bounded by trigonometry as
\begin{equation}
-1\leqslant\,({\bf a}\times{\bf a'})\cdot({\bf b'}\times{\bf b})\,\leqslant +1\,,
\end{equation}
the above inequality can be reduced to the form
\begin{equation}
-\,2\sqrt{2}\,\leqslant\,{\cal E}({\bf a},\,{\bf b})\,+\,{\cal E}({\bf a},\,{\bf b'})\,+\,
{\cal E}({\bf a'},\,{\bf b})\,-\,{\cal E}({\bf a'},\,{\bf b'})\,\leqslant+\,2\sqrt{2}\,,
\label{My-CHSH}
\end{equation}
which exhibits an extended upper bound on possible correlations. Thus, when in an EPR-Bohm experiment raw scores
${{\mathscr A}=\pm\,1}$ and ${{\mathscr B}=\pm\,1}$ are compared by coincidence counts \ocite{Aspect-1},
the normalized expectation value of their product
\begin{equation}
{\cal E}({\bf a},\,{\bf b})\,=\,\frac{\Big[C_{++}({\bf a},\,{\bf b})\,+\,C_{--}({\bf a},\,{\bf b})
\,-\,C_{+-}({\bf a},\,{\bf b})\,-\,C_{-+}({\bf a},\,{\bf b})\Big]}{\Big[C_{++}({\bf a},\,{\bf b})\,+\,C_{--}({\bf a},\,{\bf b})
\,+\,C_{+-}({\bf a},\,{\bf b})\,+\,C_{-+}({\bf a},\,{\bf b})\Big]} \label{coniii}
\end{equation}
is predicted by my model to respect, not the Bell-CHSH upper bound 2, but the Tsirel'son upper bound ${2\sqrt{2}}$, 
where ${C_{+-}({\bf a},\,{\bf b})}$ {\it etc.}${\;}$represent the number of joint occurrences of detections
${+\,1}$ along ${\bf a}$ and ${-\,1}$ along ${\bf b}$ {\it etc}.

This completes the presentation of my local, realistic, and deterministic model for the EPR-Bohm correlation.

\section{physical, mathematical, and conceptual fallacies of Weatherall's model}

\subsubsection{What is wrong with Weatherall's measurement ansatz?}

With the successful model firmly in place, we are now in a position to understand why Weatherall's model fails. To begin with,
his model is based on a different representation of the rotation group. It is in fact based, not on the
spinorial rotation group SU(2), but something akin to its tensorial cousin SO(3), which is a group of all proper rotations in
${{\rm I\!R}^3}$, {\it insensitive} to spinorial sign changes. In fact, Weatherall takes a rather odd space, namely
${{\rm I\!R}^3\wedge\,{\rm I\!R}^3}$, for the codomain of his measurement functions ${{\mathscr A}({\bf r},\,\lambda)}$, and
then introduces another projection map to arrive at the measurement results ${\{-1,\,+1\}}$. Compared to my measurement
functions (\ref{88-oi}) obtained through a smooth limiting process (\ref{amanda}), his two-step measurement process is rather
artificial. Moreover, since his is not a simply-connected, parallelized codomain such as ${S^3}$ that remains closed under
multiplication, it cannot possibly satisfy the completeness criterion of EPR \ocite{Christian}\ocite{what}.
It is therefore not surprising why
Weatherall is unable to find strong correlation among its points. Moreover, at the end of his two-step process his measurement
results ${\{-1,\,+1\}}$ are no longer the image points within the codomain
${{\rm I\!R}^3\wedge\,{\rm I\!R}^3}$ of his measurement functions. This is in sharp contrast with the situation in my model,
where my measurement results ${\{-1,\,+1\}}$ remain very much a part of the codomain ${S^3}$ of my measurement functions.
The reason why this comes about naturally within my model is because ${S^3}$, the set of unit quaternions, is a simply-connected
surface embedded in ${{\rm I\!R}^4}$ that is equipped with a {\it graded} basis made of both scalars and bivectors:
\begin{equation}
\left\{1,\;{\bf e}_2\wedge{\bf e}_3,\;{\bf e}_3\wedge{\bf e}_1,\;{\bf e}_1\wedge{\bf e}_2\right\}. \label{graded}
\end{equation}
Thus the scalars ${\{-1,\,+1\}}$ are as much a part of ${S^3}$ as the bivectors ${{\bf L}\!\left({\bf r},\,\lambda\right)}$
are, regulated by these unified basis \ocite{Christian}.
I am tempted to quip: {\it What Nature has joined together, let no man put asunder}. By contrast Weatherall's codomain is
disconnected between the space ${{\rm I\!R}^3\wedge\,{\rm I\!R}^3}$ of ``bivectors'' and the set ${\{-1,\,+1\}}$ of scalars.
His image points can thus be at best either bivectors or scalars, but not both. It is a disjoint world, more like the world
of quantum mechanics.

\begin{figure}
\hrule
\scalebox{1}{
\begin{pspicture}(4.5,-1.0)(5.0,5.9)
\psset{xunit=0.5mm,yunit=4cm}
\psaxes[axesstyle=frame,tickstyle=full,ticksize=0pt,dx=90\psxunit,Dx=180,dy=1\psyunit,Dy=+2,Oy=-1](0,0)(180,1.0)
\psline[linewidth=0.3mm,arrowinset=0.3,arrowsize=2pt 3,arrowlength=2]{->}(0,0.5)(190,0.5)
\psline[linewidth=0.2mm]{-}(45,0)(45,1)
\psline[linewidth=0.2mm]{-}(90,0)(90,1)
\psline[linewidth=0.2mm]{-}(135,0)(135,1)
\psline[linewidth=0.35mm,arrowinset=0.3,arrowsize=2pt 3,arrowlength=2]{->}(0,0)(0,1.2)
\psline[linewidth=0.35mm,linestyle=dashed,linecolor=red]{-}(0,0)(90,1)
\psline[linewidth=0.35mm,linestyle=dashed,linecolor=red]{-}(90,1)(180,0)
\put(2.1,-0.38){${90}$}
\put(6.5,-0.38){${270}$}
\put(-0.63,3.92){${+}$}
\put(-0.6,5.0){{\large ${\cal E}$}${({\bf a},\,{\bf b})}$}
\put(-0.38,1.93){${0}$}
\put(9.65,1.85){\large ${\eta_{{\bf a}{\bf b}}}$}
\psplot[linewidth=0.35mm,linecolor=blue]{0.0}{180}{x dup cos exch cos mul 1.0 mul neg 1 add}
\end{pspicture}}
\hrule
\caption{Local-realistic correlations among the points of a parallelized 3-sphere can be stronger-than-classical but not quantum.}
\vspace{0.3cm}
\label{fig-55}
\hrule
\end{figure}
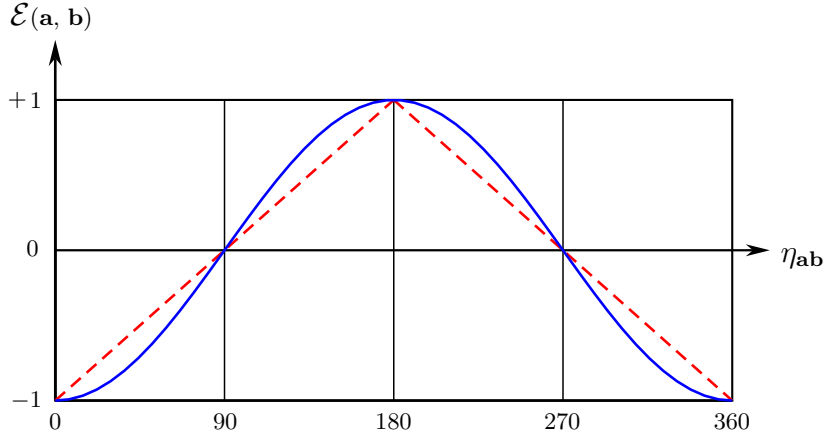

Let us, however, be more charitable to Weatherall. Let us grant him the codomain of his measurement functions to be the
connected real projective space ${{\rm I\!R}{\rm P}^3}$, which is homeomorphic to the rotation group SO(3). After all,
he does mention the Lie algebra ${so(3)}$ in one of his footnotes. So let us grant him the smooth one-step measurement process
\begin{equation}
\pm\,1\,=\,{\mathscr A}({\bf r},\,\lambda):
{\rm I\!R}^3\!\times\Lambda\longrightarrow {\rm I\!R}{\rm P}^3\sim{\rm SO}(3) \label{map-1-3}
\end{equation}
to reach the image subset ${\{-1,\,+1\}}$. This smooth map is well-defined within my model, since ${{\rm I\!R}{\rm P}^3}$ is simply
the set ${S^3}$ of unit quaternions (cf. Eq.${\,}$(\ref{nonoonpara})) with each point identified with its antipodal point
\ocite{Chris-macro}. The measurement results ${\pm\,1\in{\rm I\!R}{\rm P}^3\sim{\rm SO}(3)}$
are thus limiting points of a quaternion, just as in equation (\ref{amanda}). The map that takes us
from ${S^3}$ to ${{\rm I\!R}{\rm P}^3}$ can now be used to
project the metric ${\delta_{\mu\nu}}$ on ${S^3}$ onto ${{\rm I\!R}{\rm P}^3}$ to obtain the following induced metric on
${{\rm I\!R}{\rm P}^3}$:
\begin{align}
-\,J_{\mu\nu}\,a_{\mu}\,b_{\nu}\,=\,
\begin{cases}
-\,1\,+\,\frac{2}{\pi}\,\eta_{{\bf a}{\bf b}}
\;\;\;{\rm if} &\!\! 0 \leq \eta_{{\bf a}{\bf b}} \leq \pi \\
\\
+\,3\,-\,\frac{2}{\pi}\,\eta_{{\bf a}{\bf b}}
\;\;\;{\rm if} &\!\! \pi \leq \eta_{{\bf a}{\bf b}} \leq 2\pi\,. \label{equawhichr-2-a}
\end{cases}
\end{align}
Further details of how this metric is obtained from the metric on ${S^3}$ can be found in section III of
Ref.${\,}$\ocite{Chris-macro}. The two metrics ${\delta_{\mu\nu}}$ and ${J_{\mu\nu}}$ thus provide relative measures of
geodesic distances on the manifolds ${S^3}$ and ${{\rm I\!R}{\rm P}^3}$, respectively.
Substituting the metric on ${{\rm I\!R}{\rm P}^3}$ into
equation (\ref{corwedid}) the correlation between the points of SO(3) then works out to be
\begin{align}
{\cal E}({\bf a},\,{\bf b})\,=\,-\,J_{\mu\nu}\,a_{\mu}\,b_{\nu}\,=\,
\begin{cases}
-\,1\,+\,\frac{2}{\pi}\,\eta_{{\bf a}{\bf b}}
\;\;\;{\rm if} &\!\! 0 \leq \eta_{{\bf a}{\bf b}} \leq \pi \\
\\
+\,3\,-\,\frac{2}{\pi}\,\eta_{{\bf a}{\bf b}}
\;\;\;{\rm if} &\!\! \pi \leq \eta_{{\bf a}{\bf b}} \leq 2\pi\,. \label{equawhichr-2}
\end{cases}
\end{align}
The two sets of correlations, (\ref{robisanan}) and (\ref{equawhichr-2}), are compared in Fig.${\,}$\ref{fig-55}.
The general correlation function ${{\cal E}({\bf a},\,{\bf b})}$ derived in equation (\ref{corwedid}) can thus
serve to distinguish the geodesic distances ${{\mathscr D}({\bf a},\,{\bf b})}$ on the groups SU(2) and SO(3)
\ocite{Chris-macro}.

\subsubsection{Why did we lose the strong correlation for SO(3)?}

It is crucial to appreciate that even when we choose SO(3) as the codomain of the function ${{\mathscr A}({\bf r},\,\lambda)}$
the correct statistical procedure that must be followed is the one described in subsections II${\,}$C and II${\,}$D above. This
is because, as Weatherall himself notes, the Lie algebras of SU(2) and SO(3) are isomorphic to each other. In other words, the
local algebraic or tangent space structures on SU(2) and SO(3) are identical, but not their metrical structures in the sense of
geodesic distances. Thus the above statistical procedure, tailored to the graded basis (\ref{graded}), leading up to the general
expression (\ref{darcor}) for correlations and beyond, is equally inevitable in the case of SO(3). Comparing the two sets of
correlations resulting from this procedure---one for the
prescription (\ref{map-1-2}) and other for the prescription (\ref{map-1-3})---it is then
easy to see why we have lost the strong correlation in the second case. We started out with ${S^3}$ as a codomain of
${{\mathscr A}({\bf r},\,\lambda)}$ and then, for the case of SO(3), we identified each point of ${S^3}$ with its antipodal point.
But in doing so we lost the following spinorial rotation symmetry satisfied by ${{\bf q}(\psi,\,{\bf r})}$,
as described in equations (\ref{clacct}) and (\ref{non-clacct}) above:
\begin{equation}
{\bf q}(\psi+2\kappa\pi,\,{\bf r})\,=\,-\,{\bf q}(\psi,\,{\bf r})\,\;\;\text{for}\;\,\kappa=1,3,5,7,\dots
\end{equation}
In other words, by identifying the antipodal points of ${S^3}$ we lost the sensitivity to spinorial sign changes.
As a result, ${{\bf q}(\psi,\,{\bf r})}$ now represents the state of a rotating body that returns to itself after
any and all multiples of ${2\pi}$ rotation:
\begin{equation}
{\bf q}(\psi+2\kappa\pi,\,{\bf r})\,=\,+\,{\bf q}(\psi,\,{\bf r})\,\;\;\text{for}\;\,\text{any}\;\,\kappa=0,1,2,3,\dots
\end{equation}
This is the real reason why we lost the strong correlation for the SO(3) case. The reason Weatherall has argued for
is an artifact of his bad choice of measurement functions. It stems from a failure to appreciate the unified nature
of the graded basis (\ref{graded}) and the associated fact that the scalars ${\{-1,\,+1\}}$ and the bivectors
${{\bf L}\!\left({\bf r},\,\lambda\right)}$ occur as image points within the same codomain ${S^3}$ in my model.
Thus the loss of correlation has nothing to do with the fact that ultimately the measurement functions must map to the
image subset ${\{-1,\,+1\}}$. They {\it manifestly do} in my model [cf. Eqs.${\,}$(\ref{88-oi}), (\ref{99-oi}), and
(\ref{amanda})]. The raw scalar numbers ${{\mathscr A}=\pm\,1}$ and ${{\mathscr B}=\pm\,1}$ mapped to the image subset
${\{-1,\,+1\}}$---according to my model---are indeed the numbers used by Alice and Bob for calculating the correlation
in the usual manner. And when, at the end of their experiment, they evaluate the statistical quantity
${{\cal S}({\bf a},\,{\bf a'},\,{\bf b},\,{\bf b'})}$ involving these numbers as
\begin{equation}
{\cal S}({\bf a},\,{\bf a'},\,{\bf b},\,{\bf b'})\,=\,\lim_{\,n\,\gg\,1}\Bigg[\frac{1}{n}\sum_{k\,=\,1}^{n}\,\big\{
{\mathscr A}_{\bf a}({\lambda}^k)\,{\mathscr B}_{\bf b}({\lambda}^k)\,+\,
{\mathscr A}_{\bf a}({\lambda}^k)\,{\mathscr B}_{\bf b'}({\lambda}^k)\,+\,
{\mathscr A}_{\bf a'}({\lambda}^k)\,{\mathscr B}_{\bf b}({\lambda}^k)\,-\,
{\mathscr A}_{\bf a'}({\lambda}^k)\,{\mathscr B}_{\bf b'}({\lambda}^k)\big\}\Bigg],
\end{equation}
they will inevitably find that it exceeds the bound of 2 and extends to the bound of ${2\sqrt{2}}$.
This conclusion may seem odd from the perspective of Bell-type reasoning,
but the evidence presented for it in subsection II${\,}$F is incontrovertible.

\section{Concluding Remarks}

In addition to the main issue discussed above, it is instructive to reflect on the broader reasons
why Weatherall's model fails. We can in fact identify at least six erroneous steps which
engender the failure of his model from the start:
\begin{enumerate}
\item[${\bf 1.}$] Choice of incomplete codomain of the measurement functions ${{\mathscr A}({\bf r},\,\lambda)}$
(although with correct image points ${\pm\,1}$).
\item[${\bf 2.}$] Neglect of putting scalars and bivectors on equal footing within a single, comprehensive, real number system.
\item[${\bf 3.}$] Failure to implement a spinor representation of SU(2) by recognizing the significance of spinorial sign
changes.
\item[${\bf 4.}$] Lack of appreciation of the role played by the parallelizing torsion within ${S^3}$ for the
existence and strength of strong correlations (or not recognizing the discipline of parallelization as the true
source of strong correlations).
\item[${\bf 5.}$] Failure to appreciate how errors propagate within ${S^3}$, when taken as a codomain of the measurement functions.
\item[${\bf 6.}$] Neglect of the correct statistical procedure in the derivations of both
the correlation and the upper bound ${2\sqrt{2}}$.
\end{enumerate}
Although interconnected, any one of these reasons is sufficient for the failure of Weatherall's model. Recognizing this,
I must conclude that, contrary to first impressions, Weatherall's thinly veiled criticism of my work is entirely vacuous.

\vspace{-0.15cm}

\acknowledgments
I am grateful to Lucien Hardy for several months of correspondence which led to improvements in section II${\,}$D.
I am also grateful to Martin Castell for his hospitality in the Materials Department of the University of Oxford.
This work was funded by a grant from the Foundational Questions Institute (FQXi) Fund, a donor advised fund of the
Silicon Valley Community Foundation on the basis of proposal FQXi-MGA-1215 to the Foundational Questions Institute.

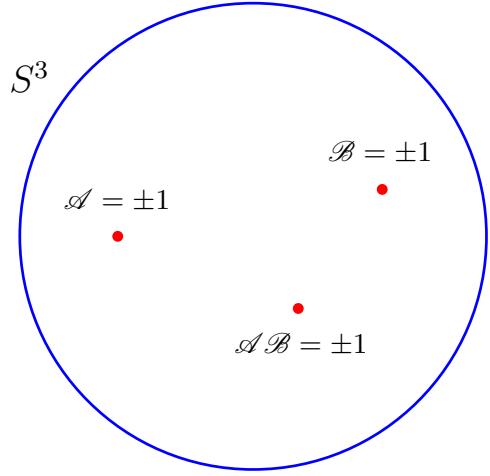
\begin{figure}
\hrule
\scalebox{1.2}{
\begin{pspicture}(-5.9,-3.7)(2.5,2.8)

\pscircle[linewidth=0.3mm,linecolor=blue](-1.8,-0.45){2.6}

\put(-4.5,1.15){{\large ${S^3}$}}

\put(-2.01,-1.75){{${{\mathscr A}{\mathscr B}=\pm1}$}}

\put(-3.9,-0.15){{${{\mathscr A}=\pm1}$}}

\put(-0.97,0.38){{${{\mathscr B}=\pm1}$}}

\pscircle[fillcolor=red,linecolor=red,fillstyle=solid](-1.3,-1.25){0.06}

\pscircle[fillcolor=red,linecolor=red,fillstyle=solid](-3.3,-0.45){0.06}

\pscircle[fillcolor=red,linecolor=red,fillstyle=solid](-0.37,0.07){0.06}

\end{pspicture}}
\hrule
\caption{The results ${\mathscr A}$ and ${\mathscr B}$ observed by Alice and Bob are points of a parallelized
3-sphere, ${S^3}$. Since ${S^3}$ remains closed under multiplication, the product ${{\mathscr A}{\mathscr B}}$
is also a point of the same 3-sphere, with its value, ${\pm1}$, dictated by the topology of ${S^3}$.}
\vspace{0.3cm}
\label{fig-56}
\hrule
\end{figure}

\appendix
\section{${\;\;\,}$Local-realistic Violations of not only the Bell-CHSH Inequality, but also the Clauser-Horne Inequality}

\subsection{Local-realistic violations of the Bell-CHSH inequality}

In this appendix I demonstrate explicit violations of both the CHSH and the Clauser-Horne inequalities within the strictly
local model described in the previous pages. This will further lay bare the vacuity of Weatherall's analysis. To this end,
let us first confirm the following probabilistic predictions of quantum mechanics within our local model:
\begin{align}
P^{++}:=\,P\left\{{\mathscr A}=+1,\;{\mathscr B}=+1;\;{\mathscr A}{\mathscr B}=+1
\;|\;\eta_{{\bf a}{\bf b}}\,\right\}\,&=\,\frac{1}{2}
\sin^2\left(\frac{\eta_{{\bf a}{\bf b}}}{2}\right), \\
P^{--}:=\,P\left\{{\mathscr A}=-1,\;{\mathscr B}=-1;\;{\mathscr A}{\mathscr B}=+1
\;|\;\eta_{{\bf a}{\bf b}}\,\right\}\,&=\,\frac{1}{2}
\sin^2\left(\frac{\eta_{{\bf a}{\bf b}}}{2}\right), \\
P^{-+}:=\,P\left\{{\mathscr A}=-1,\;{\mathscr B}=+1;\;{\mathscr A}{\mathscr B}=-1
\;|\;\eta_{{\bf a}{\bf b}}\,\right\}\,&=\,\frac{1}{2}
\cos^2\left(\frac{\eta_{{\bf a}{\bf b}}}{2}\right), \\
\text{and}\;\;\;\;P^{+-}:=\,P\left\{{\mathscr A}=+1,\;{\mathscr B}=-1;\;{\mathscr A}{\mathscr B}=-1
\;|\;\eta_{{\bf a}{\bf b}}\,\right\}\,&=\,\frac{1}{2}
\cos^2\left(\frac{\eta_{{\bf a}{\bf b}}}{2}\right).
\end{align}
Here ${\eta_{{\bf a}{\bf b}}}$ is the angle between ${\bf a}$ and ${\bf b}$. As is well known, confirming these probabilities
is equivalent to confirming
\begin{align}
{\cal E}({\bf a},\,{\bf b})\,&=\lim_{\,n\,\gg\,1}\left[\frac{1}{n}\sum_{i\,=\,1}^{n}\,
{\mathscr A}({\bf a},\,\lambda^i)\;{\mathscr B}({\bf b},\,\lambda^i)\right] \notag \\
&=\,\frac{(P^{++}\,+\,P^{--})\times({\mathscr A}{\mathscr B}=+1)
\,+\,(P^{-+}\,+\,P^{+-})\times({\mathscr A}{\mathscr B}=-1)}{P^{++}\,+\,P^{--}\,+\,P^{-+}\,+\,P^{+-}}\, \notag \\
&=\,(P^{++}\,+\,P^{--}\,-\,P^{-+}\,-\,P^{+-})/(P^{++}\,+\,P^{--}\,+\,P^{-+}\,+\,P^{+-})\, \notag \\
&=\,\frac{1}{2}\sin^2\left(\frac{\eta_{{\bf a}{\bf b}}}{2}\right)\,
+\,\frac{1}{2}\sin^2\left(\frac{\eta_{{\bf a}{\bf b}}}{2}\right)\,
-\,\frac{1}{2}\cos^2\left(\frac{\eta_{{\bf a}{\bf b}}}{2}\right)\,
-\,\frac{1}{2}\cos^2\left(\frac{\eta_{{\bf a}{\bf b}}}{2}\right)\,=\,-\,\cos\eta_{{\bf a}{\bf b}}\,, \label{corappenir}
\end{align}
which is the strong correlation (\ref{robisanan}) predicted by our local model (see also Ref.${\,}$\ocite{Aspect-1}). To confirm
this result explicitly, suppose initially a pair of spins, ${-{\bf L}({\bf e}_o,\,\lambda)}$ and ${{\bf L}({\bf e}_o,\,\lambda)}$,
is created in a singlet configuration, where ${{\bf e}_o}$ is a random direction about which the spins are rotating at the moment
of their creation, and ${\lambda=\pm\,1}$ is their initial state, which, in our model, is taken to be the orientation
of a parallelized 3-sphere within which the events ${\mathscr A}$ and ${\mathscr B}$ are occurring.
At a later time, ${t}$, let Alice and Bob detect the spins with
detectors ${{\bf D}({\bf a})}$ and ${{\bf D}({\bf b})}$, respectively. Thus, at time ${t}$, the spin bivectors
${-{\bf L}({\bf e}_t,\,\lambda)}$ and ${{\bf L}({\bf e}_t,\,\lambda)}$ are projected by them onto the detector bivectors
${{\bf D}({\bf a})}$ and ${{\bf D}({\bf b})}$, respectively. Then, as we defined in equations (\ref{88-oi}) and (\ref{99-oi})
above, we have the following measurement results [cf. also Eq.${\,}$(\ref{amanda})]:
\begin{align}
S^3\ni\pm\,1\,=\,{\mathscr A}({\bf a},\,\lambda)\,
&=\,\lim_{{\bf e}_o\rightarrow\,{\bf a}}\,{\mathscr A}({\bf a},\,{\bf e}_o,\,\lambda) \notag \\
&=\,\lim_{{\bf e}_o\rightarrow\,{\bf a}}\,\left\{\,-\,{\bf D}({\bf a})\,{\bf L}({\bf e}_o,\,\lambda)\,\right\}\, \notag \\
&=\,\lim_{{\bf e}_o\rightarrow\,{\bf a}}\,\left\{\,(\,-\,I\cdot{\bf a})(\,\lambda\,I\cdot{\bf e}_o)\,\right\}
=\,\lim_{{\bf e}_o\rightarrow\,{\bf a}}\,\left\{\,+\lambda\,{\bf a}\cdot{\bf e}_o
                                   \,+\,\lambda\,I\cdot({\bf a}\times{\bf e}_o)\,\right\} \label{888-oi}
\end{align}
and
\begin{align}
S^3\ni\pm\,1\,=\,{\mathscr B}({\bf b},\,\lambda)\,
&=\,\lim_{{\bf e}_o\rightarrow\,{\bf b}}\,{\mathscr B}({\bf b},\,{\bf e}_o,\,\lambda) \notag \\
&=\,\lim_{{\bf e}_o\rightarrow\,{\bf b}}\,\left\{\,+\,{\bf D}({\bf b})\,{\bf L}({\bf e}_o,\,\lambda)\,\right\}\, \notag \\
&=\,\lim_{{\bf e}_o\rightarrow\,{\bf b}}\,\left\{\,(\,+\,I\cdot{\bf b})(\,\lambda\,I\cdot{\bf e}_o)\,\right\}
=\,\lim_{{\bf e}_o\rightarrow\,{\bf b}}\,\left\{\,-\lambda\,{\bf b}\cdot{\bf e}_o
                                   \,-\,\lambda\,I\cdot({\bf b}\times{\bf e}_o)\,\right\}\!. \label{999-oi}
\end{align}
Let me stress once again that the measurement results ${{\mathscr A}({\bf a},\,{\lambda})}$ and
${{\mathscr B}({\bf b},\,{\lambda})}$ as defined above, in addition to being manifestly realistic, are strictly local and
deterministically determined numbers within the 3-sphere. In fact, they are not even contextual. Alice's measurement result
${{\mathscr A}({\bf a},\,{\lambda})}$---although it refers to a freely chosen direction ${\bf a}$---depends only on the initial
state ${\lambda}$; and likewise, Bob's measurement result ${{\mathscr B}({\bf b},\,{\lambda})}$---although it refers to a freely
chosen direction ${\bf b}$---depends only on the initial state ${\lambda}$. In other words, they are statistically independent
events within ${S^3}$.

Now, for the purposes of our calculations below, it is convenient to make a note of the following useful identities: 
\begin{equation}
{\bf L}({\bf e}_o,\,\lambda)\,=\,\lambda\,{\bf D}({\bf e}_o)\,=\,\pm\,{\bf D}({\bf e}_o)
\,=\,(\pm\,I)\cdot{\bf e}_o\,=\,I\cdot(\pm\,{\bf e}_o)\,=\,{\bf D}(\pm\,{\bf e}_o)\,\,=\,{\bf D}(\lambda\,{\bf e}_o)\,
\;\;\;\text{and}\;\;\;{\bf L}^2({\bf e}_o)\,=\,{\bf D}^2({\bf e}_o)\,=\,-1.
\end{equation}
Thus, with ${{\bf e}_o}$ uniformly distributed in ${{\rm I\!R}^3}$, the measurement results (\ref{888-oi}) and (\ref{999-oi})
can be expressed equivalently as
\begin{align}
S^3\ni\pm\,1\,=\,{\mathscr A}({\bf a};\,{\bf e}_o,\,\lambda)\,=\,\,-\,{\bf D}({\bf a})\,{\bf D}(\lambda\,{\bf e}_o)\,=\,
\begin{cases}
+\,1\;\;\;\;\;{\rm if} &\lambda\,{\bf e}_o\,=\,+\,{\bf a} \\
-\,1\;\;\;\;\;{\rm if} &\lambda\,{\bf e}_o\,=\,-\,{\bf a}
\end{cases} \label{88-oij}
\end{align}
and
\begin{align}
S^3\ni\pm\,1\,=\,{\mathscr B}({\bf b};\,{\bf e}_o,\,\lambda)\,=\,\,+\,{\bf D}({\bf b})\,{\bf D}(\lambda\,{\bf e}_o)\,=\,
\begin{cases}
-\,1\;\;\;\;\;{\rm if} &\lambda\,{\bf e}_o\,=\,+\,{\bf b} \\
+\,1\;\;\;\;\;{\rm if} &\lambda\,{\bf e}_o\,=\,-\,{\bf b}\,.
\end{cases} \label{99-oij}
\end{align}
Note again that measurement functions ${{\mathscr A}({\bf a};\,{\bf e}_o,\,\lambda)}$ and
${{\mathscr B}({\bf b};\,{\bf e}_o,\,\lambda)}$ are manifestly local, precisely as demanded by Bell.
Apart from the {\it common causes} ${\lambda}$ and ${{\bf e}_o}$, their values depend {\it only} on
${\bf a}$ and ${\bf b}$, respectively, and nothing else. Moreover, since ${{\bf e}_o}$ is uniformly distributed
within ${{\rm I\!R}^3}$ and ${\lambda}$ has 50/50 chance of being ${\pm\,1}$, their averages vanish:
\begin{equation}
\overline{{\mathscr A}({\bf a};\,{\bf e}_o,\,\lambda)}\,=\,0\,=\,
\overline{{\mathscr B}({\bf b};\,{\bf e}_o,\,\lambda)}\,.\label{avecon}
\end{equation}

\begin{figure}
\hrule
\vspace{-3.4cm}
\scalebox{2.8}{
\begin{pspicture}(-0.11,0.0)(5.0,5.0)

\epsfig{figure=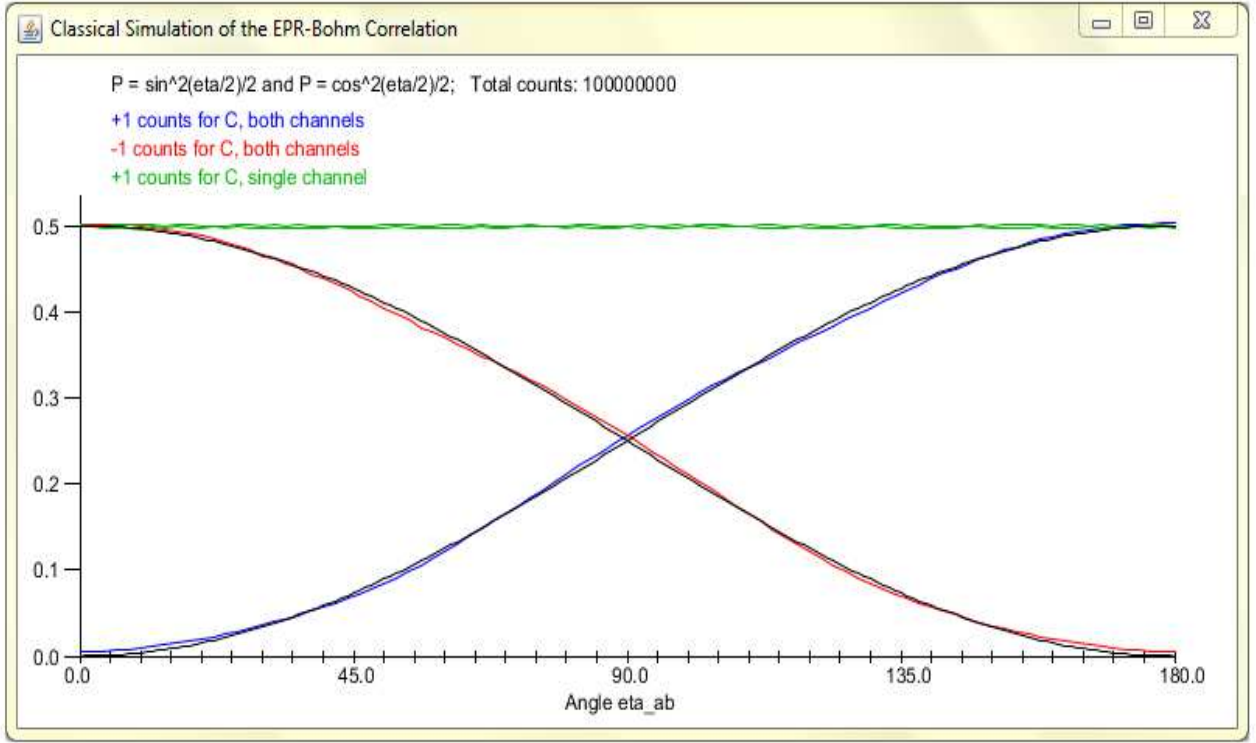,height=102pt,width=169.8pt}

\end{pspicture}}
\vspace{0.2cm}
\hrule
\caption{A simulation of outcome probabilities for simultaneously occurring measurement events ${{\mathscr A}=\pm1}$ and
${{\mathscr B}=\pm1}$ within a parallelized 3-sphere. The probabilistic predictions of the manifestly local model described
in the text match exactly with the corresponding predictions of quantum mechanics \ocite{Chris-macro}. 
The x-axis in the chart specifies the
values of the relative angle ${\eta_{{\bf a}{\bf b}}}$, and the y-axis records the rate of simultaneous occurrences of events
${{\mathscr A}=\pm1}$ and ${{\mathscr B}=\pm1}$ as a function of ${\eta_{{\bf a}{\bf b}}}$. Thus, for example, the blue
curve depicts the ratio of the number of simultaneous occurrences of events ${{\mathscr A}=+1}$ and ${{\mathscr B}=+1}$
(or ${{\mathscr A}=-1}$ and ${{\mathscr B}=-1}$) over the total number of occurrences and non-occurrences of such events.
The straight line at y = 0.5, on the other hand, depicts the number of occurrences of events such as ${{\mathscr A}=+1}$
at one station (with no detector present at the other station) over the total number of occurrences and non-occurrences
of such events. The computer code for the simulation is available at https://github.com/chenopodium/JCS.
Further discussion on the relevant issues can be found also on my blog at http://libertesphilosophica.info/blog/. I am grateful
to {\sf Chantal Roth} for kindly writing the code and for insightful discussions.}
\label{fig-57}
\vspace{0.3cm}
\hrule
\end{figure}

Let us confirm this averaging conditions in some more detail by explicitly
rewriting equations (\ref{888-oi}) and (\ref{999-oi}) as
\begin{align}
S^3\ni\pm\,1\,&=\,{\mathscr A}({\bf a};\,{\bf e}_o,\,\lambda)\,
=\,\lim_{\eta_{{\bf a}\lambda{\bf e}_o}\,\rightarrow\,\kappa\pi}\,\left\{\,+\,\cos(\,\eta_{{\bf a}\lambda{\bf e}_o}\,)
\,+\,I\cdot{\bf c}_{\bf a}(\lambda{\bf e}_o)\,\sin(\,\eta_{{\bf a}\lambda{\bf e}_o}\,)\right\} \label{888999-oi} \\
\text{and}\;\;\;\;S^3\ni\pm\,1\,&=\,{\mathscr B}({\bf b};\,{\bf e}_o,\,\lambda)\,
=\,\lim_{\eta_{{\bf b}\lambda{\bf e}_o}\,\rightarrow\,\kappa\pi}\,\left\{\,-\,\cos(\,\eta_{{\bf b}\lambda{\bf e}_o}\,)
\,-\,I\cdot{\bf c}_{\bf b}(\lambda{\bf e}_o)\,\sin(\,\eta_{{\bf b}\lambda{\bf e}_o}\,)\right\}\!. \label{999888-oi}
\end{align}
Here ${\kappa=0,\,1,\,2,\,\dots}$; ${\eta_{{\bf a}\lambda{\bf e}_o}}$ is the angle between ${\bf a}$ and ${\lambda\,{\bf e}_o}$;
${{\bf c}_{\bf a}({\bf e}_o):={\bf a}\times{\bf e}_o/|{\bf a}\times{\bf e}_o|}$; and
${{\bf c}_{\bf b}({\bf e}_o):={\bf b}\times{\bf e}_o/|{\bf b}\times{\bf e}_o|}$. Now, since ${\lambda{\bf e}_o}$ is a {\it random}
unit vector in ${{\rm I\!R}^3}$, the quantity
${{\cal C}({\bf a};\,{\bf e}_o,\,\lambda):=\cos(\,\eta_{{\bf a}\lambda{\bf e}_o})}$, being a function of
this random variable, acts as a loaded die, taking values in the interval ${[-1,\,+1]}$, with 50\% chance
of landing on its face marked${\;+1}$:
\begin{equation}
-1\,\leq\,{\cal C}({\bf a};\,{\bf e}_o,\,\lambda)\,\leq\,+1\,,  \label{notpr}
\end{equation}
with 50\% chance for the occurrence\footnote{To understand these odds, let
${p({\bf a};\,{\bf e}_o,\,\lambda)=|\cos(\,\eta_{{\bf a}\lambda{\bf e}_o})|}$ be the probability density function
defined over a 2-sphere of unit radius with respect to a fixed reference vector ${{\bf a}={\bf x}}$, for observing a measurement
result ${{\mathscr A}=+1}$ or ${{\mathscr A}=-1}$:
\begin{equation}
p({\bf a};\,{\bf e}_o,\,\lambda):S^2\rightarrow [0,\,1]\,. \notag
\end{equation}
For the purposes of our calculation a particular vector
${\lambda{\bf e}_o}$ on the 2-sphere can be specified by the angle ${\eta_{{\bf a}\lambda{\bf e}_o}}$ in the range
${[\,0,\,\pi\,]}$ with respect to the fixed vector ${\bf a}$. Consequently, we can calculate the probability
${P({\mathscr A},\,{\bf a})}$ of occurrence of a result ${{\mathscr A}=+1}$ or ${-1}$ by integrating over ${S^2}$ as follows:
\begin{align}
P({\mathscr A},\,{\bf a})\,=\,\frac{1}{4\pi} \int_{S^2} \,p({\bf a};\,{\bf e}_o,\,\lambda) \;\,d\Omega\,
=\,\frac{1}{4\pi}\int_0^{\pi}\int_0^{2\pi} |\cos(\,\eta_{{\bf a}\lambda{\bf e}_o})|\;\,\sin(\,\eta_{{\bf a}\lambda{\bf e}_o})
\,d\xi_{{\bf y}\lambda{\bf e}_o}d\eta_{{\bf a}\lambda{\bf e}_o}\,, \notag
\end{align}
where ${d\Omega=\sin(\,\eta_{{\bf a}\lambda{\bf e}_o})\,d\xi_{{\bf y}\lambda{\bf e}_o}\,d\eta_{{\bf a}\lambda{\bf e}_o}}$
is the differential solid angle on the 2-sphere, with ${\xi_{{\bf y}\lambda{\bf e}_o}}$ being the azimuthal
angle from the ${y}$-axis and ${\eta_{{\bf a}\lambda{\bf e}_o}}$ being the polar angle from ${{\bf a}={\bf x}}$.
Now it is evident from
Eq.${\,}$(\ref{888999-oi}) that for the result ${{\mathscr A}=+1}$ the angle ${\eta_{{\bf a}\lambda{\bf e}_o}}$ must fall
in the range ${[\,0,\,\pi/2\,]}$ for ${\lambda=+1}$, and in the range ${[\,\pi/2,\,\pi\,]}$ for ${\lambda=-1}$.
Consequently, the probability of observing ${{\mathscr A}=+1}$ is
\begin{align}
P({\mathscr A}=+1,\,{\bf a})\,
&=\,\frac{1}{4\pi}\int_0^{\frac{\pi}{2}}
\int_0^{2\pi} \cos(\,\eta_{{\bf a}{\bf e}_o})\;\,\sin(\,\eta_{{\bf a}{\bf e}_o})
\,d\xi_{{\bf y}{\bf e}_o}d\eta_{{\bf a}{\bf e}_o}\, \notag \\
&\;\;\;\;\;\;\;\;\;\;\;\;\;\;\;\;\;\;\;\;\;\;+
\,\frac{1}{4\pi}\int_{\frac{\pi}{2}}^{\pi}
\int_0^{2\pi} \cos(\,\pi-\eta_{{\bf a}{\bf e}_o})\;\,\sin(\,\pi-\eta_{{\bf a}{\bf e}_o})
\,d\xi_{{\bf y}{\bf e}_o}d\eta_{{\bf a}{\bf e}_o} \notag \\
&=\,\frac{1}{2}\int_0^{\frac{\pi}{2}}\cos(\,\eta_{{\bf a}{\bf e}_o})\,\sin(\,\eta_{{\bf a}{\bf e}_o})\;\,d\eta_{{\bf a}{\bf e}_o}
\,-\,\frac{1}{2}\int_{\frac{\pi}{2}}^{\pi} \cos(\,\eta_{{\bf a}{\bf e}_o})\,\sin(\,\eta_{{\bf a}{\bf e}_o})
\;\,d\eta_{{\bf a}{\bf e}_o}\, \notag \\
&=\,\frac{1}{2}\,\times\,\frac{\,\sin^2(\,\eta_{{\bf a}{\bf e}_o})}{2}\Bigg|_0^{\frac{\pi}{2}}\,-\;
\frac{1}{2}\,\times\,\frac{\,\sin^2(\,\eta_{{\bf a}{\bf e}_o})}{2}\Bigg|_{\frac{\pi}{2}}^{\pi}
\,=\,\frac{1}{4}\,+\,\frac{1}{4}\,=\,\frac{1}{2}\,. \notag 
\end{align}
Similarly, it is evident from Eq.${\,}$(\ref{888999-oi}) that for the result ${{\mathscr A}=-1}$ the
angle ${\eta_{{\bf a}\lambda{\bf e}_o}}$ must fall in the range ${[\,\pi/2,\,\pi\,]}$ for ${\lambda=+1}$, and
in the range ${[\,0,\,\pi/2\,]}$ for ${\lambda=-1}$, with the probability density remaining positive:
${0\leq p({\bf a};\,{\bf e}_o,\,\lambda)\leq 1}$. Consequently, the probability of observing ${{\mathscr A}=-1}$ is
\begin{align}
P({\mathscr A}=-1,\,{\bf a})\,
&=\,-\,\frac{1}{4\pi}\int_0^{\frac{\pi}{2}}
\int_0^{2\pi} \cos(\,\pi-\eta_{{\bf a}{\bf e}_o})\;\,\sin(\,\pi-\eta_{{\bf a}{\bf e}_o})
\,d\xi_{{\bf y}{\bf e}_o}d\eta_{{\bf a}{\bf e}_o}\, \notag \\
&\;\;\;\;\;\;\;\;\;\;\;\;\;\;\;\;\;\;\;\;\;\;-
\,\frac{1}{4\pi}\int_{\frac{\pi}{2}}^{\pi}
\int_0^{2\pi} \cos(\,\eta_{{\bf a}{\bf e}_o})\;\,\sin(\,\eta_{{\bf a}{\bf e}_o})
\,d\xi_{{\bf y}{\bf e}_o}d\eta_{{\bf a}{\bf e}_o} \notag \\
&=\,\frac{1}{2}\int_0^{\frac{\pi}{2}}\cos(\,\eta_{{\bf a}{\bf e}_o})\,\sin(\,\eta_{{\bf a}{\bf e}_o})\;\,d\eta_{{\bf a}{\bf e}_o}
\,-\,\frac{1}{2}\int_{\frac{\pi}{2}}^{\pi} \cos(\,\eta_{{\bf a}{\bf e}_o})\,\sin(\,\eta_{{\bf a}{\bf e}_o})
\;\,d\eta_{{\bf a}{\bf e}_o}\, \notag \\
&=\,\frac{1}{2}\,\times\,\frac{\,\sin^2(\,\eta_{{\bf a}{\bf e}_o})}{2}\Bigg|_0^{\frac{\pi}{2}}\,-\;
\frac{1}{2}\,\times\,\frac{\,\sin^2(\,\eta_{{\bf a}{\bf e}_o})}{2}\Bigg|_{\frac{\pi}{2}}^{\pi}
\,=\,\frac{1}{4}\,+\,\frac{1}{4}\,=\,\frac{1}{2}\,. \notag
\end{align}
These probabilities [and the corresponding odds for ${{\cal C}({\bf a};\,{\bf e}_o,\,\lambda)}$]
are confirmed in the event-by-event simulation described in Fig.${\,}$\ref{fig-57}.}
\begin{equation}
{\cal C}({\bf a};\,{\bf e}_o,\,\lambda)\,=\,+1\,, \notag \label{notpr-2}
\end{equation}
and similarly for ${{\cal C}({\bf a};\,{\bf e}_o,\,\lambda)=-1}$. Note that for a fixed orientation ${\lambda}$, say ${\lambda=+1}$,
${{\cal C}({\bf a};\,{\bf e}_o,\,+1)}$ has only 25\% chance of landing on its face marked ${+1}$. But for
${\lambda=+1}$ {\it or} ${\lambda=-1}$, which is equivalent to the possibilities ${\eta_{{\bf a}{\bf e}_o}}$ {\it or}
${\,\pi-\eta_{{\bf a}{\bf e}_o}}$ in the ${\cos(\,\eta_{{\bf a}\lambda{\bf e}_o})}$ part of
${{\mathscr A}({\bf a};\,{\bf e}_o,\,\lambda)}$,
${{\cal C}({\bf a};\,{\bf e}_o,\,\lambda)}$ has further 25\% chance of landing on its face ${+1}$, because
${\lambda=+1}$ {\it or} ${\lambda=-1}$ are two mutually exclusive possibilities. Consequently,
the die ${{\cal C}({\bf a};\,{\bf e}_o,\,\lambda)}$ has exactly 50/50 chance of landing on its faces ${+1}$ or ${-1}$ in
general. Thus, the measurement outcomes ${{\mathscr A}=+1}$ or ${-1}$ observed by Alice (regardless of Bob)
is a result of a throw of the die ${{\cal C}({\bf a};\,{\bf e}_o,\,\lambda)}$, which has 50/50 chance of landing on its face
marked ${+1}$ or ${-1}$, depending on the values of ${\eta_{{\bf a}\lambda{\bf e}_o}}$. One can confirm these odds for
${{\cal C}({\bf a};\,{\bf e}_o,\,\lambda)}$ in a computer simulation, as in Fig.${\,}$\ref{fig-57}.

The question now is: How do the actual values ${\mathscr A}$ and ${\mathscr B}$ of the functions
${{\mathscr A}({\bf a};\,{\bf e}_o,\,\lambda)}$ and ${{\mathscr B}({\bf b};\,{\bf e}_o,\,\lambda)}$ come about within
the 3-sphere when they are observed {\it simultaneously} by Alice and Bob? To answer this question, recall that 3-sphere remains
closed under multiplication, with its points represented by quaternions of the form (\ref{defi-2}),
and ${\mathscr A}$ and ${\mathscr B}$
are limiting values of such quaternions. Therefore the values ${\mathscr A}$ and ${\mathscr B}$ observed by Alice and Bob
are constrained by the value ${{\mathscr A}{\mathscr B}({\bf a},\,{\bf b};\,{\bf e}_o)}$ of the product
${{\mathscr A}({\bf a};\,{\bf e}_o,\,\lambda){\mathscr B}({\bf b};\,{\bf e}_o,\,\lambda)}$, which is a limiting value of
the {\it product} quaternion:
\begin{equation}
S^3\ni\pm\,1\,=\,{\mathscr A}({\bf a};\,{\bf e}_o,\,\lambda){\mathscr B}({\bf b};\,{\bf e}_o,\,\lambda)\,=
\,\lim_{\substack{{\bf e}_o\rightarrow\,\pm{\bf a} \\ {\bf e}_o\rightarrow\,\pm{\bf b}}}\,
\left[\{\,+\,{\bf a}\cdot{\bf e}_o\,+\,I\cdot({\bf a}\times{\bf e}_o)\,\}\,
\{\,-\,{\bf b}\cdot{\bf e}_o\,-\,I\cdot({\bf b}\times{\bf e}_o)\}\right],\label{ntt-2}
\end{equation}
where I have omitted explicit reference to ${\lambda}$ in the RHS because it drops out from the product.
Let me stress again that, since ${S^3}$ remains closed under multiplication, a limit like this one is the
only legitimate way of determining the correct values of ${\mathscr A}$, ${\mathscr B}$, and ${{\mathscr A}{\mathscr B}}$,
since all three of them are constrained to be what they possibly could be by the very geometry and topology of the
3-sphere. In other words, by taking the orientation ${\lambda}$ of ${S^3}$ as the initial state of the spin system, we have
in fact taken the entire geometry and topology of ${S^3(\lambda)}$ as the {\it common cause} that determines the actual
values ${\mathscr A}$ and ${\mathscr B}$, which are thus dictated to be what they turn out to be by the product
${{\mathscr A}{\mathscr B}}$ resulting in the limit (\ref{ntt-2}). Hence our goal is to find the correct product quaternion
whose limit is the value ${{\mathscr A}{\mathscr B}=+1}$ or ${-1}$.

To this end, let us work out the product appearing on the RHS of the above limit equation explicitly, which gives
\begin{align}
\{\,+\,&{\bf a}\cdot{\bf e}_o\,+\,I\cdot({\bf a}\times{\bf e}_o)\,\}\,
\{\,-\,{\bf b}\cdot{\bf e}_o\,-\,I\cdot({\bf b}\times{\bf e}_o)\} \notag \\
&\;\;\;\;\,=\,-\,({\bf a}\cdot{\bf e}_o)({\bf b}\cdot{\bf e}_o)\,+\,({\bf a}\times{\bf e}_o)\cdot({\bf b}\times{\bf e}_o)\,-\,
I\cdot\left\{\,({\bf a}\cdot{\bf e}_o)({\bf b}\times{\bf e}_o)\,+\,
({\bf b}\cdot{\bf e}_o)({\bf a}\times{\bf e}_o)\,-\,
({\bf a}\times{\bf e}_o)\times({\bf b}\times{\bf e}_o)\,\right\} \notag \\
&\;\;\;\;\;\;\;\;\;
\longrightarrow\,{\mathscr A}{\mathscr B}({\bf a},\,{\bf b};\,{\bf e}_o)\,=\,\pm\,1\in S^3.\label{mni-identity}
\end{align}
Although it may not be obvious at first sight, this product is necessarily a unit quaternion of the form (\ref{defi-2}):
\begin{equation}
{\bf q}(\psi,\,{\bf r})\,=\,\cos\frac{\psi}{2}\,+\,(I\cdot{\bf r})\,\sin\frac{\psi}{2}
\,\Longleftrightarrow\,{\cal C}({\bf a},\,{\bf b};\,{\bf e}_o)\,+\,
I\cdot{\bf c}_{\bf k}({\bf e}_o)\;\,{\cal S}({\bf a},\,{\bf b};\,{\bf e}_o)\,,\;\;\;\text{with}\;\;\;
{\cal C}^2+\,{\cal S}^2\,=\,1\,. \label{defi-3}
\end{equation}
In other words, the product (\ref{mni-identity}) is a sum of a scalar part and a bivector part, and when its scalar part,
${\cal C}$, reduces to ${\pm1}$, its bivector part reduces to zero,
because the coefficient, ${\cal S}$, of the bivector ${I\cdot{\bf c}_{\bf k}({\bf e}_o)}$
reduces to zero. In view of (\ref{ntt-2}), (\ref{mni-identity}), and (\ref{defi-3}) it is thus clear that the limits
${{\mathscr A}{\mathscr B}({\bf a},\,{\bf b};\,{\bf e}_o)\,\rightarrow\,+1}$ or ${-1}$ are equivalent to the limits
\begin{align}
{\cal C}({\bf a},\,{\bf b};\,{\bf e}_o)\,:=\,
-\,({\bf a}\cdot{\bf e}_o)({\bf b}\cdot{\bf e}_o)\,+\,({\bf a}\times{\bf e}_o)\cdot({\bf b}\times{\bf e}_o)\,
&\longrightarrow\,+\,1 \label{c-1}\\
\text{or}\;\;\;\;\;
{\cal C}({\bf a},\,{\bf b};\,{\bf e}_o)\,:=\,
-\,({\bf a}\cdot{\bf e}_o)({\bf b}\cdot{\bf e}_o)\,+\,({\bf a}\times{\bf e}_o)\cdot({\bf b}\times{\bf e}_o)\,
&\longrightarrow\,-\,1\,. \label{c-2}
\end{align}
But since ${{\bf e}_o}$ is a uniformly distributed {\it random} unit vector in ${{\rm I\!R}^3}$, the quantity
${{\cal C}({\bf a},\,{\bf b};\,{\bf e}_o)}$ is also a {\it random} variable, taking values from the interval ${[-1,\,+1]}$:
\begin{equation}
-1\,\leq\,{\cal C}({\bf a},\,{\bf b};\,{\bf e}_o)\,\leq\,+1\,.
\end{equation}
Thus, just as ${{\cal C}({\bf a};\,{\bf e}_o,\,\lambda)}$ in Eq.${\,}$(\ref{notpr}), ${{\cal C}({\bf a},\,{\bf b};\,{\bf e}_o)}$
is a die with certain propensities for landing on its faces ${+1}$ or ${-1}$. In particular, for a sufficiently large number
of random vectors ${{\bf e}_o}$, the random variable ${{\cal C}({\bf a},\,{\bf b};\,{\bf e}_o)}$ has 47\% chance of landing
on its face marked ${-1}$ for ${{\bf b}=+{\bf a}}$, and 47\% chance of landing on its face marked ${+1}$ for ${{\bf b}=-{\bf a}}$.
What is more, from the LHS of the conditions (\ref{c-1}) and (\ref{c-2}), using the limits ${{\bf e}_o\rightarrow\,{\bf a}}$
and ${{\bf e}_o\rightarrow\,{\bf b}\,}$, it is easy to verify that
\begin{align}
{\mathscr A}({\bf a};\,{\bf e}_o,\,\lambda){\mathscr B}({\bf b};\,{\bf e}_o,\,\lambda)
\,=\,{\mathscr A}{\mathscr B}({\bf a},\,{\bf b};\,{\bf e}_o)\,
&\longrightarrow\,-1\;\;\;\text{when}\;\;\;{\bf b=+a}, \\
\text{whereas}\;\;\;\;\;
{\mathscr A}({\bf a};\,{\bf e}_o,\,\lambda){\mathscr B}({\bf b};\,{\bf e}_o,\,\lambda)
\,=\,{\mathscr A}{\mathscr B}({\bf a},\,{\bf b};\,{\bf e}_o)\,
&\longrightarrow\,+1\;\;\;\text{when}\;\;\;{\bf b=-a}\,,
\end{align}
where the second result is due to the spinorial sign change that the quaternion
${{\mathscr A}{\mathscr B}({\bf a},\,{\bf b=-a})}$ undergoes with respect to the quaternion
${{\mathscr A}{\mathscr B}({\bf a},\,{\bf b=+a})}$ upon rotation by angle ${\eta_{{\bf a}{\bf b}}=\pi}$.
[${\,}$It is worth noting here that Weatherall's model is incapable of exhibiting such a change
from ${{\mathscr A}{\mathscr B}=-1}$ to ${{\mathscr A}{\mathscr B}=+1}$ in the product of the observed numbers
${\mathscr A}$ and ${\mathscr B}$. His model has been deliberately manufactured to fail
this first test, in anticipation of his intended goal.]

The above results are of course two very special cases. They are the cases when the measurement axes chosen by Alice
and Bob happen to be aligned or anti-aligned with each other. In more general cases, when the measurement axes of Alice
and Bob happen to subtend angles different from ${\eta_{{\bf a}{\bf b}}=0}$ or ${\pi}$, it takes a bit more effort to
deduce what values the results ${\mathscr A}$ and ${\mathscr B}$ and their product ${{\mathscr A}{\mathscr B}}$ within
${S^3}$ would turn out to have. In fact, because ${{\cal C}({\bf a},\,{\bf b};\,{\bf e}_o)}$ is a random variable,
in more general cases statistical considerations are inevitable. The question we must ask in the general cases is the
following: Given the random variable ${{\cal C}({\bf a},\,{\bf b};\,{\bf e}_o)}$ taking values in the interval
${[-1,\,+1]}$, what are the chances---as functions of the angle ${\eta_{{\bf a}{\bf b}}}$ between ${\bf a}$ and
${\bf b}$---for the occurrences of the events ${{\mathscr A}{\mathscr B}=+1}$ and ${{\mathscr A}{\mathscr B}=-1}$?
The answer to this question can be obtained by computing the following set of conditional probabilities:
\begin{align}
&P\left\{\,{\mathscr A}=+1,\;{\mathscr B}=+1;\;{\mathscr A}{\mathscr B}=+1\;|\;\eta_{{\bf a}{\bf b}}\,\right\}\,=\,
P\left\{\,{\cal C}({\bf a},\,{\bf b};\,{\bf e}_o,\,\phi_o^p,\,\phi_o^q,\,\phi_o^r,\,\phi_o^s)
=+1\;|\;\eta_{{\bf a}{\bf b}}\,\right\}, \label{ghp-1}\\
&P\left\{\,{\mathscr A}=-1,\;{\mathscr B}=-1;\;{\mathscr A}{\mathscr B}=+1\;|\;\eta_{{\bf a}{\bf b}}\,\right\}\,=\,
P\left\{\,{\cal C}({\bf a},\,{\bf b};\,{\bf e}_o,\,\phi_o^p,\,\phi_o^q,\,\phi_o^r,\,\phi_o^s)
=+1\;|\;\eta_{{\bf a}{\bf b}}\,\right\}, \\
&P\left\{\,{\mathscr A}=+1,\;{\mathscr B}=-1;\;{\mathscr A}{\mathscr B}=-1\;|\;\eta_{{\bf a}{\bf b}}\,\right\}\,=\,
P\left\{\,{\cal C}({\bf a},\,{\bf b};\,{\bf e}_o,\,\phi_o^p,\,\phi_o^q,\,\phi_o^r,\,\phi_o^s)
=-1\;|\;\eta_{{\bf a}{\bf b}}\,\right\}, \\
\text{and}\;\;\;\;
&P\left\{\,{\mathscr A}=-1,\;{\mathscr B}=+1;\;{\mathscr A}{\mathscr B}=-1\;|\;\eta_{{\bf a}{\bf b}}\,\right\}\,=\,
P\left\{\,{\cal C}({\bf a},\,{\bf b};\,{\bf e}_o,\,\phi_o^p,\,\phi_o^q,\,\phi_o^r,\,\phi_o^s)
=-1\;|\;\eta_{{\bf a}{\bf b}}\,\right\}, \label{ghp-4}
\end{align}
where the random variable ${{\cal C}({\bf a},\,{\bf b};\,{\bf e}_o)}$---which (as we saw above) can be thought of as a loaded
die---is now given by
\begin{equation}
{\cal C}({\bf a},\,{\bf b};\,{\bf e}_o,\,\phi_o^p,\,\phi_o^q,\,\phi_o^r,\,\phi_o^s)=
\left\{\,-\,\cos(\eta_{{\bf a}{\bf e}_o}\,+\,\phi_o^p)\,\cos(\eta_{{\bf b}{\bf e}_o}\,+\,\phi_o^r)
\,+\,\cos\eta_{{\bf c}_{\bf a}{\bf c}_{\bf b}}\,
\sin(\eta_{{\bf a}{\bf e}_o}\,+\,\phi_o^q)\,\sin(\eta_{{\bf b}{\bf e}_o}\,+\,\phi_o^s)\right\}\!/N_{\bf a}N_{\bf b}.
\label{c-23-want}
\end{equation}
In the form of Eq.${\,}$(\ref{defi-3}), this is the scalar part of the product of the following two manifestly local quaternions:
\begin{align}
{\bf q}({\bf a};\,{\bf e}_o,\,\lambda,\,\phi_o^p,\,\phi_o^q)\,
&=\,\left\{\,+\,\cos(\,\eta_{{\bf a}\lambda{\bf e}_o}\,+\,\phi_o^p\,)
\,+\,I\cdot{\bf c}_{\bf a}(\lambda{\bf e}_o)\,\sin(\,\eta_{{\bf a}\lambda{\bf e}_o}\,+\,\phi_o^q\,)\,\right\}\!/N_{\bf a}
\,\equiv\,\left\{\,{\cal C}_{\bf a}+(I\cdot{\bf c}_{\bf a})\,{\cal S}_{\bf a}\,\right\} \label{quat-aaa}\\
\text{and}\;\;
{\bf q}({\bf b};\,{\bf e}_o,\,\lambda,\,\phi_o^r,\,\phi_o^s)\,
&=\,\left\{\,-\,\cos(\,\eta_{{\bf b}\lambda{\bf e}_o}\,+\,\phi_o^r\,)
\,-\,I\cdot{\bf c}_{\bf b}(\lambda{\bf e}_o)\,\sin(\,\eta_{{\bf b}\lambda{\bf e}_o}\,+\,\phi_o^s\,)\,\right\}\!/N_{\bf b}
\,\equiv\,\left\{\,{\cal C}_{\bf b}+(I\cdot{\bf c}_{\bf b})\,{\cal S}_{\bf b}\right\}\!, \label{quat-bbb}
\end{align}
where ${N_{\bf a}=\sqrt{\cos^2(\eta_{{\bf a}{\bf e}_o}+\,\phi_o^p)+\sin^2(\eta_{{\bf a}{\bf e}_o}+\,\phi_o^q)}\,}$
and ${N_{\bf b}=\sqrt{\cos^2(\eta_{{\bf b}{\bf e}_o}+\,\phi_o^r)+\sin^2(\eta_{{\bf b}{\bf e}_o}+\,\phi_o^s)}}$ are
the normalizing factors. Needless to say,
${{\cal C}({\bf a},\,{\bf b};\,{\bf e}_o,\,\phi_o^p,\,\phi_o^q,\,\phi_o^r,\,\phi_o^s)}$ specified in Eq.${\,}$(\ref{c-23-want})
is the same as that specified in Eqs.${\,}$(\ref{c-1}) and (\ref{c-2}) apart from the phase shifts ${\phi_o^p}$, ${\phi_o^q}$,
${\phi_o^r}$, and ${\phi_o^s}$. These phase shifts contribute to the rotations of the product quaternion about the random vector
${{\bf c}_{\bf k}({\bf e}_o)}$ [which is different in general from the vectors ${{\bf c}_{\bf a}(\lambda{\bf e}_o)}$ and
${{\bf c}_{\bf b}(\lambda{\bf e}_o)}$]. They are simply numerical constants, independent of the parameter vectors ${\bf a}$ and
${\bf b}$, or of the random vector ${{\bf e}_o}$. They are thus parts of the geometry of the
3-sphere. Together with ${{\bf e}_o}$, they form what Bell referred to as the ``past causes.''

In terms of the product of the quaternions (\ref{quat-aaa}) and (\ref{quat-bbb}), we are now in a position to express
the measurement results ${{\mathscr A}({\bf a};\,{\bf e}_o,\,\lambda)}$ and ${{\mathscr B}({\bf b};\,{\bf e}_o,\,\lambda)}$,
and their product ${{\mathscr A}{\mathscr B}({\bf a},\,{\bf b};\,{\bf e}_o)}$, in the following manifestly local-realistic forms: 
\begin{align}
S^3\ni\pm\,1\,=\,{\mathscr A}({\bf a};\,{\bf e}_o,\,\lambda)
\,&=\lim_{{\cal C}_{\bf a}\,\rightarrow\,\pm1}\left\{\,{\cal C}({\bf a};\,{\bf e}_o,\,\lambda,\,\phi_o^p,\,\phi_o^q)
\,+\,I\cdot{\bf c}_{\bf a}(\lambda{\bf e}_o)\,\;{\cal S}({\bf a};\,{\bf e}_o,\,\lambda,\,\phi_o^p,\,\phi_o^q)
\right\}\!, \label{eieipo} \\
S^3\ni\pm\,1\,=\,{\mathscr B}({\bf b};\,{\bf e}_o,\,\lambda)
\,&=\lim_{{\cal C}_{\bf b}\,\rightarrow\,\pm1}\left\{\,{\cal C}({\bf b};\,{\bf e}_o,\,\lambda,\,\phi_o^r,\,\phi_o^s)
\,+\,I\cdot{\bf c}_{\bf b}(\lambda{\bf e}_o)\,\;{\cal S}({\bf b};\,{\bf e}_o,\,\lambda,\,\phi_o^r,\,\phi_o^s)
\right\}\!,\;\;\;\text{and} \label{eieilo} \\
S^3\ni\pm\,1\,=\,{\mathscr A}{\mathscr B}({\bf a},\,{\bf b};\,{\bf e}_o)\,
\,&=\lim_{{\cal C}_{\bf ab}\,\rightarrow\,\pm1}
\left\{\,{\cal C}({\bf a},\,{\bf b};\,{\bf e}_o,\,\phi_o^p,\,\phi_o^q,\,\phi_o^r,\,\phi_o^s)
\,+\,I\cdot{\bf c}_{\bf k}({\bf e}_o)\,\;{\cal S}({\bf a},\,{\bf b};\,{\bf e}_o,\,\phi_o^p,\,\phi_o^q,\,\phi_o^r,\,\phi_o^s)
\right\}\!. \label{eieioo}
\end{align}
Here the vector ${{\bf c}_{\bf k}({\bf e}_o)}$ depends on both ${{\bf c}_{\bf a}(\lambda{\bf e}_o)}$ and
${{\bf c}_{\bf b}(\lambda{\bf e}_o)}$, as well as on the other past causes. Although we do not require its explicit expression,
it can be easily worked out by evaluating the
geometric product of the quaternions ${{\bf q}({\bf a};\,{\bf e}_o,\,\lambda,\,\phi_o^p,\,\phi_o^q)}$ and
${{\bf q}({\bf b};\,{\bf e}_o,\,\lambda,\,\phi_o^r,\,\phi_o^s)}$.
The 3-sphere analogue of Bell's factorizability condition then appears as
\begin{align}
\left\{\,{\cal C}_{\bf ab}+(I\cdot{\bf c}_{\bf k})\,{\cal S}_{\bf ab}\,\right\}\,&=\;
\left\{\,{\cal C}_{\bf a}+(I\cdot{\bf c}_{\bf a})\,{\cal S}_{\bf a}\,\right\}\;
\left\{\,{\cal C}_{\bf b}+(I\cdot{\bf c}_{\bf b})\,{\cal S}_{\bf b}\,\right\}
\;\;\;\;(\,\text{with all}\;\;{\cal C}^2+\,{\cal S}^2=\,1\,), \notag \\
\text{\it i.e.,}\;\;\;S^3\ni{\bf q}_{{\bf a}{\bf b}}\,&=\,{\bf q}_{\bf a}\,{\bf q}_{\bf b}
\;\;\;\text{for}\;\,{\bf q}_{\bf a}\;\text{and}\;\,{\bf q}_{\bf b}\in S^3, \label{nimpo-1}
\end{align}
which, in the measurement limits
(\ref{eieipo}) and (\ref{eieilo}), reduces to the factorization of the corresponding scalar values:
\begin{equation}
S^3\ni\pm\,1\,=\,\boxed{{\mathscr A}{\mathscr B}({\bf a},\,{\bf b};\,{\bf e}_o)
\,=\,{\mathscr A}({\bf a};\,{\bf e}_o,\,\lambda)\times{\mathscr B}({\bf b};\,{\bf e}_o,\,\lambda)}
\,=\,\left\{\pm\,1\right\}\times\left\{\pm\,1\right\}\,=\,\pm\,1\in S^3. \label{nimpo-2}
\end{equation}
Needless to say that the constraints (\ref{nimpo-1}) and (\ref{nimpo-2}), which are manifestations of the geometry and topology
of the 3-sphere, put strong restrictions on the possible values of the phase shifts ${\phi_o^p}$, ${\phi_o^q}$, ${\phi_o^r}$, and
${\phi_o^s}$. One possible set of their values in degrees is ${\{\,0,\;0,\;-86.92,\;\text{and}\;37.99\,\}}$. With these values,
the scalar parts of ${{\mathscr A}({\bf a};\,{\bf e}_o,\,\lambda)}$ and ${{\mathscr B}({\bf b};\,{\bf e}_o,\,\lambda)}$ are given
by ${\;{\cal C}({\bf a};\,{\bf e}_o,\,\lambda,\,\phi_o^p,\,\phi_o^q)
=\left\{\,+\,\cos(\eta_{{\bf a}\lambda{\bf e}_o}+\,\phi_o^p)\right\}\!/N_{\bf a}\;}$
and ${\;{\cal C}({\bf b};\,{\bf e}_o,\,\lambda,\,\phi_o^r,\,\phi_o^s)=
\left\{\,-\,\cos(\eta_{{\bf b}\lambda{\bf e}_o}+\,\phi_o^r)\right\}\!/N_{\bf b}\,}$,\break respectively, whereas the scalar part
${{\cal C}({\bf a},\,{\bf b};\,{\bf e}_o,\,\phi_o^p,\,\phi_o^q,\,\phi_o^r,\,\phi_o^s)}$ of
${{\mathscr A}{\mathscr B}({\bf a},\,{\bf b};\,{\bf e}_o)}$ is the one stated in equation (\ref{c-23-want}).

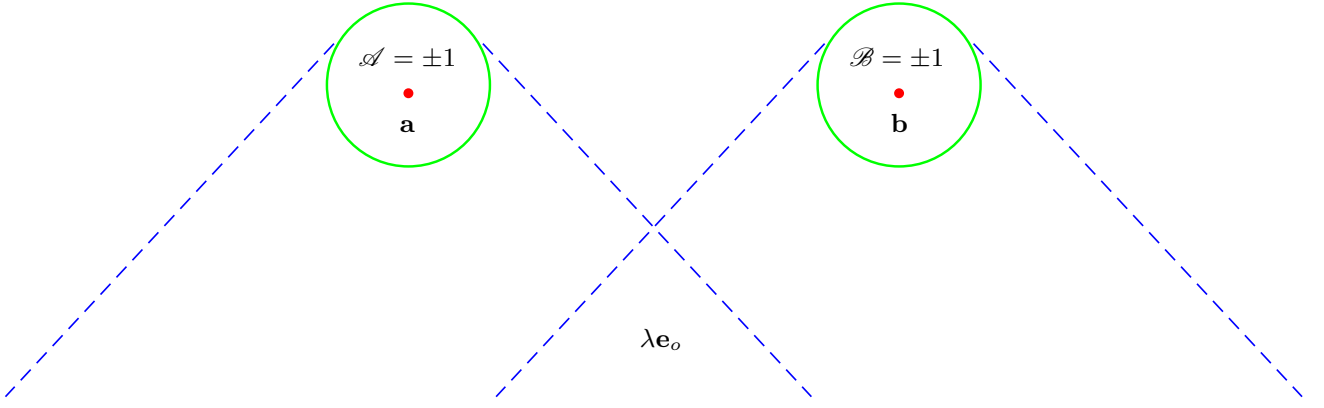
\begin{figure}
\hrule
\scalebox{1.1}{
\begin{pspicture}(-5.9,-3.9)(2.5,2.8)

\put(-5.4,1.0){{${{\mathscr A}=\pm1}$}}

\put(-4.9,0.2){{${\bf a}$}}

\put(-2.0,-2.4){{${\lambda{\bf e}_o}$}}

\put(0.53,1.0){{${{\mathscr B}=\pm1}$}}

\put(1.03,0.2){{${\bf b}$}}

\pscircle[fillcolor=red,linecolor=red,fillstyle=solid](-4.8,0.67){0.06}

\pscircle[fillcolor=red,linecolor=red,fillstyle=solid](1.13,0.67){0.06}

\pscircle[linewidth=0.3mm,linecolor=green](-4.8,0.77){1.0}

\pscircle[linewidth=0.3mm,linecolor=green](1.13,0.77){1.0}

\psline[linewidth=0.2mm,linestyle=dashed,linecolor=blue]{-}(-3.9,1.27)(0.07,-3.0)

\psline[linewidth=0.2mm,linestyle=dashed,linecolor=blue]{-}(-5.7,1.27)(-9.67,-3.0)

\psline[linewidth=0.2mm,linestyle=dashed,linecolor=blue]{-}(2.03,1.27)(6.0,-3.0)

\psline[linewidth=0.2mm,linestyle=dashed,linecolor=blue]{-}(0.23,1.27)(-3.74,-3.0)

\end{pspicture}}
\hrule
\caption{The measurement results ${{\mathscr A}({\bf a};\,{\bf e}_o,\,\lambda)}$ and ${{\mathscr B}({\bf b};\,{\bf e}_o,\,\lambda)}$
observed independently by Alice and Bob are manifestly local.}
\vspace{0.3cm}
\label{fig-60}
\hrule
\end{figure}

\begin{figure}
\hrule
\vspace{-2.8cm}
\scalebox{2.2}{
\begin{pspicture}(-1.0,0.1)(5.0,5.1)

\epsfig{figure=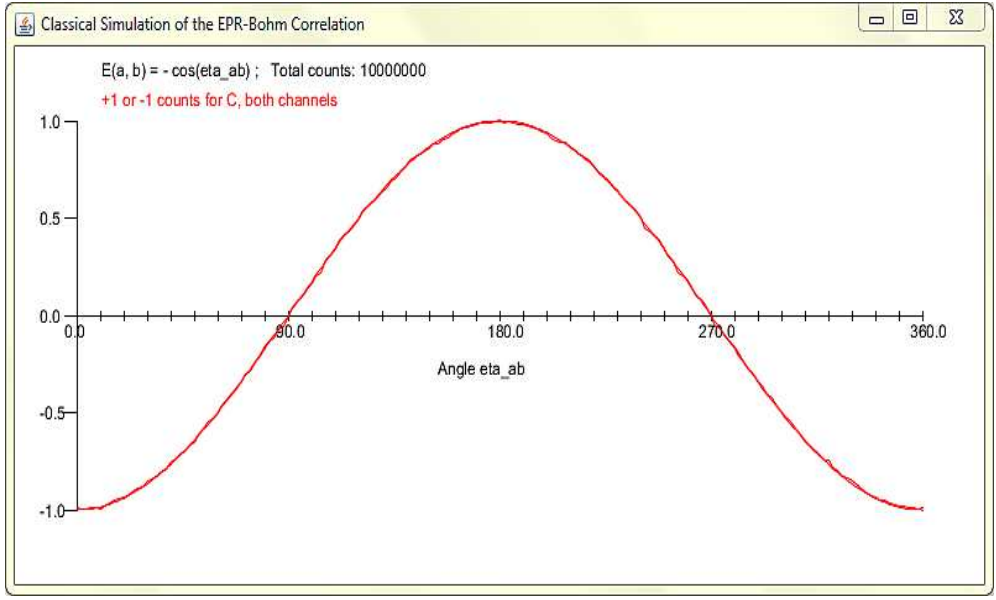,height=102pt,width=169.8pt}

\end{pspicture}}
\vspace{0.2cm}
\hrule
\caption{A simulation of the correlation between simultaneously occurring measurement events ${{\mathscr A}=\pm1}$ and
${{\mathscr B}=\pm1}$ within a parallelized 3-sphere. As we noted in the caption of Fig.${\,}$\ref{fig-57}, the probabilistic
predictions of the manifestly local model described in the text match exactly with the corresponding predictions of quantum
mechanics \ocite{Christian}\ocite{Chris-macro}\ocite{origins}.
The correlation generated in this simulation is essentially an ``addition'' of the coincident
probabilities given in Eqs.${\,}$(\ref{lesson}) to (\ref{lesson-6}), as computed in Eq.${\,}$(\ref{molnot}).}
\label{fig-59}
\vspace{0.3cm}
\hrule
\end{figure}

As depicted in Fig.${\,}$\ref{fig-56}, (\ref{eieipo}), (\ref{eieilo}), and (\ref{eieioo}) are the explicit expressions of the
three points of the 3-sphere we have been looking for. Using these expressions we can now compute the probabilities for the
occurrences of simultaneous events at the two ends of the EPR experiment:
${P\{{\mathscr A}=+1,\;{\mathscr B}=+1\}}$, ${P\{{\mathscr A}=-1,\;{\mathscr B}=-1\}}$, ${P\{{\mathscr A}=-1,\;{\mathscr B}=+1\}}$,
and ${P\{{\mathscr A}=+1,\;{\mathscr B}=-1\}}$. From the factorizability conditions (\ref{nimpo-1}) and (\ref{nimpo-2}) it is
easy to see, however, that in the measurement limits these probabilities are causally constrained by the respective values of
${{\cal C}({\bf a},\,{\bf b};\,{\bf e}_o,\,\phi_o^p,\,\phi_o^q,\,\phi_o^r,\,\phi_o^s)}$, as indicated in Eqs.${\,}$(\ref{ghp-1})
to (\ref{ghp-4}). Thus, for example, the probability of the simultaneous occurrence of the results
${{\mathscr A}=-1}$ and ${{\mathscr B}=+1}$ is given by the probability of the die 
${{\cal C}({\bf a},\,{\bf b};\,{\bf e}_o,\,\phi_o^p,\,\phi_o^q,\,\phi_o^r,\,\phi_o^s)}$ landing on its face marked ${-1}$.
The results of numerical computations of such probabilities are illustrated in Fig.${\,}$\ref{fig-57}. They turn out to be
as follows:
\begin{align}
P_1^{+}({\bf a})&=P\left\{\,{\cal C}({\bf a};\,{\bf e}_o,\,\lambda,\,\phi_o^p,\,\phi_o^q)
=+1\;|\;\text{no detector present at {\bf b}}\,\right\}\,=\,\frac{1}{2}\,, \label{lesson-1}\\
P_2^{-}({\bf b})&=P\left\{\,{\cal C}({\bf b};\,{\bf e}_o,\,\lambda,\,\phi_o^r,\,\phi_o^s)
=-1\;|\;\text{no detector present at {\bf a}}\,\right\}\,=\,\frac{1}{2}\,, \label{lesson-2}\\
P_{12}^{++}&=\,P\left\{\,{\cal C}({\bf a},\,{\bf b};\,{\bf e}_o,\,\phi_o^p,\,\phi_o^q,\,\phi_o^r,\,\phi_o^s)
=+1\;|\;\eta_{{\bf a}{\bf b}}\,\right\}\,=\,\frac{1}{2}
\sin^2\left(\frac{\eta_{{\bf a}{\bf b}}}{2}\right), \label{lesson}\\
P_{12}^{--}&=\,P\left\{\,{\cal C}({\bf a},\,{\bf b};\,{\bf e}_o,\,\phi_o^p,\,\phi_o^q,\,\phi_o^r,\,\phi_o^s)
=+1\;|\;\eta_{{\bf a}{\bf b}}\,\right\}\,=\,\frac{1}{2}
\sin^2\left(\frac{\eta_{{\bf a}{\bf b}}}{2}\right), \\
P_{12}^{-+}&=\,P\left\{\,{\cal C}({\bf a},\,{\bf b};\,{\bf e}_o,\,\phi_o^p,\,\phi_o^q,\,\phi_o^r,\,\phi_o^s)
=-1\;|\;\eta_{{\bf a}{\bf b}}\,\right\}\,=\,\frac{1}{2}
\cos^2\left(\frac{\eta_{{\bf a}{\bf b}}}{2}\right), \\
\text{and}\;\;\;\;P_{12}^{+-}&=\,P\left\{\,{\cal C}({\bf a},\,{\bf b};\,{\bf e}_o,\,\phi_o^p,\,\phi_o^q,\,\phi_o^r,\,\phi_o^s)
=-1\;|\;\eta_{{\bf a}{\bf b}}\,\right\}\,=\,\frac{1}{2}
\cos^2\left(\frac{\eta_{{\bf a}{\bf b}}}{2}\right).  \label{lesson-6}
\end{align}
As described in Eq.${\,}$(\ref{corappenir}), the correlation predicted by our local model can now be readily calculated as follows: 
\begin{align}
{\cal E}({\bf a},\,{\bf b})\,=\,\frac{P_{12}^{++}\,+\,P_{12}^{--}\,-\,P_{12}^{-+}\,-\,P_{12}^{+-}}{P_{12}^{++}\,+\,P_{12}^{--}\,+\,P_{12}^{-+}\,+\,P_{12}^{+-}}\,
&=\,\frac{1}{2}\sin^2\left(\frac{\eta_{{\bf a}{\bf b}}}{2}\right)\,
+\,\frac{1}{2}\sin^2\left(\frac{\eta_{{\bf a}{\bf b}}}{2}\right)\,
-\,\frac{1}{2}\cos^2\left(\frac{\eta_{{\bf a}{\bf b}}}{2}\right)\,
-\,\frac{1}{2}\cos^2\left(\frac{\eta_{{\bf a}{\bf b}}}{2}\right)\, \notag \\
&=\,\sin^2\left(\frac{\eta_{{\bf a}{\bf b}}}{2}\right)\,
-\,\cos^2\left(\frac{\eta_{{\bf a}{\bf b}}}{2}\right)\, \notag \\
&=\,-\,\cos\eta_{{\bf a}{\bf b}}\,. \label{molnot}
\end{align}
This correlation can also be computed directly in the simulation described in Fig.${\,}$\ref{fig-57}.
The result is depicted in Fig.${\,}$\ref{fig-59}.
 
Let me stress that the above
correlation is produced entirely within a strictly local and deterministic model of the physical reality.
As we have already noted several times,
the measurement results ${{\mathscr A}({\bf a};\,{\bf e}_o,\,\lambda)}$ and ${{\mathscr B}({\bf b};\,{\bf e}_o,\,\lambda)}$
defined in (\ref{eieipo}) and (\ref{eieilo}) are manifestly local. 
Moreover, the product quaternions ${{\bf q}_{{\bf a}{\bf b}}}$, and hence their scalar limits
${{\mathscr A}({\bf a};\,{\bf b};\,{\bf e}_o,\,\lambda)}$, are intrinsically factorizable, as shown in
equations (\ref{nimpo-1}) and (\ref{nimpo-2}).
${{\mathscr A}({\bf a};\,{\bf e}_o,\,\lambda)}$ and ${{\mathscr B}({\bf b};\,{\bf e}_o,\,\lambda)}$ are thus
determined by a common cause---namely, the geometry and
topology of the 3-sphere---and the local vectors ${\bf a}$ and ${\bf b}$. In practical terms, this common cause is translated into
the set ${\{\lambda{\bf e}_o,\,\phi_o^p,\,\phi_o^q,\,\phi_o^r,\,\phi_o^s\}}$. The actual measurement results
${\mathscr A}$ and ${\mathscr B}$ observed by Alice and Bob are thus only dependent on these common causes and the local vectors.

For completeness, let me also note that the above correlation gives rise to the following string of expectation values:
\begin{equation}
{\cal E}({\bf a},\,{\bf b})\,+\,{\cal E}({\bf a},\,{\bf b'})\,+\,
{\cal E}({\bf a'},\,{\bf b})\,-\,{\cal E}({\bf a'},\,{\bf b'})\,=\,
-\,\cos\eta_{{\bf a}{\bf b}}\,-\,\cos\eta_{{\bf a}{\bf b'}}\,-\,
\cos\eta_{{\bf a'}{\bf b}}\,+\,\cos\eta_{{\bf a'}{\bf b'}}\,.
\end{equation}
It is well known that the right-hand-side of the above equation violates the celebrated Bell-CHSH inequality \ocite{Clauser}.

\subsection{Local-realistic
violations of the Clauser-Horne inequality\protect\footnotemark}\footnotetext{I am grateful to Lucien Hardy for
raising the question about the violations of Clauser-Horne inequality in the present context.}

It is worth noting here that the local-realistic model described above violates not only the Bell-CHSH inequality, but also
the Clauser-Horne inequality \ocite{Report}:
\begin{equation}
-1\,\leq\,P_{12}^{++}({\bf a},\,{\bf b})\,-\,P_{12}^{++}({\bf a},\,{\bf b'})\,+\,
P_{12}^{++}({\bf a'},\,{\bf b})\,+\,P_{12}^{++}({\bf a'},\,{\bf b'})
\,-\,P_{1}^{+}({\bf a'})
\,-\,P_{2}^{+}({\bf b})\,\leq\,0\,.
\end{equation}
To appreciate this, recall the following predictions of our local model:
\begin{equation}
P_{1}^{+}({\bf a'})\,=\,P_{2}^{+}({\bf b})\,=\,\frac{1}{2}\;\;\;\;
\text{and}\;\;\;\;P_{12}^{++}({\bf a},\,{\bf b})\,=\,\frac{1}{2}\sin^2\left(\frac{\eta_{{\bf a}{\bf b}}}{2}\right). \label{pp2}
\end{equation}
These predictions are the same as the ones stated in Eqs.${\,}$(\ref{lesson-1}) to (\ref{lesson})---which, in turn, are the same
as those predicted by quantum mechanics. Consequently, Clauser-Horne inequality is inevitably violated by our local model.

It is also worth recalling here that once the topology of the codomain of the measurement functions is correctly specified, not
only the EPR-Bohm correlation, but also the correlations predicted by the rotationally non-invariant quantum states such as the
GHZ states and the Hardy state---and in fact those predicted by {\it ALL} quantum states---can be reproduced exactly
in a purely classical, local-realistic manner \ocite{what}\ocite{origins}\ocite{illusion-1}.
Thus, contrary to the widespread belief, the correlations exhibited by such
states are not irreducible quantum effects, but purely local-realistic, topological effects.
Needless to say, this vindicates Einstein's
suspicion that quantum state merely describes statistical ensemble of physical systems, and not the individual physical system.
It is this inevitable conclusion that Weatherall is resisting.

\subsection{Elegant, powerful, and succinct calculation of the correlation}

The above results once again confirm the fact that EPR-Bohm correlations are local-realistic correlations among the points of
a parallelized 3-sphere \ocite{Christian}\ocite{Chris-macro}.
As we saw in section II, however, this fact can be expressed more elegantly by
understanding how random errors propagate within a parallelized 3-sphere. In particular, we saw that EPR-Bohm correlations can
be derived by recognizing that the raw scores ${{\mathscr A}({\bf a},\,{\lambda})}$ and ${{\mathscr B}({\bf b},\,{\lambda})}$
are generated within ${S^3}$ with {\it different} bivectorial scales of dispersion, and hence the correct correlation between
them can be determined only by calculating the covariation of the corresponding standardized variables
${{\bf L}({\bf a},\,\lambda)}$ and ${{\bf L}({\bf b},\,\lambda)}$:
\begin{align}
{\cal E}({\bf a},\,{\bf b})\,=\lim_{\,n\,\gg\,1}\left[\frac{1}{n}\sum_{i\,=\,1}^{n}\,
{\mathscr A}({\bf a},\,\lambda^i)\;{\mathscr B}({\bf b},\,\lambda^i)\right]
\,=\lim_{\,n\,\gg\,1}\left[\frac{1}{n}\sum_{i\,=\,1}^{n}\,
{\bf L}({\bf a},\,\lambda^i){\bf L}({\bf b},\,\lambda^i)\right]\,=\,-\,{\bf a}\cdot{\bf b}\,, \label{afort}
\end{align}
where
\begin{equation}
{\bf L}({\bf a},\,\lambda^i)\,{\bf L}({\bf b},\,\lambda^i)\,\equiv\,-\,{\bf a}\cdot{\bf b}\,-\,
{\bf L}({\bf a}\times{\bf b},\,\lambda^i)\,\equiv\,-\,{\bf a}\cdot{\bf b}\,-\,
\lambda^i\,{\bf D}({\bf a}\times{\bf b}),
\end{equation}
and the standardized variables are defined as
\begin{equation} 
{\bf L}({\bf a},\,\lambda)\,:=
\frac{\,{\bf q}(\psi,\,{\bf a},\,\lambda)\,-\,m({\bf q})}{\sigma[\,{\bf q}(\psi,\,{\bf a},\,\lambda)]}\,=
\frac{\,{\mathscr A}({\bf a},\,\lambda)\,-\,m({\mathscr A})}{\sigma[\,{\mathscr A}({\bf a},\,\lambda)]}\,.
\end{equation}
Note that this definition holds for any point ${{\bf q}(\psi,\,{\bf a},\,\lambda)}$ of ${S^3}$,
and not just for the limiting point ${{\bf q}(\psi,\,{\bf a},\,\lambda)={\mathscr A}({\bf a},\,\lambda)}$
${=\pm1}$
obtained in equation (\ref{amanda}).
The EPR-Bohm correlations ${-\,{\bf a}\cdot{\bf b}}$ are thus correlations between any two points
${{\bf q}(\psi,\,{\bf a},\,\lambda)}$ and ${{\bf q}(\psi,\,{\bf b},\,\lambda)}$ of ${S^3}$, with the scalar
points ${{\mathscr A}({\bf a},\,\lambda)=\pm1}$ and ${{\mathscr B}({\bf b},\,\lambda)=\pm1}$
being only a special case.

It is also noteworthy that the correlation between the raw scores ${{\mathscr A}({\bf a},\,{\lambda})}$ and
${{\mathscr B}({\bf b},\,{\lambda})}$ is determined in Eq.${\,}$(\ref{afort})
by calculating their covariance divided by the product
of their standard deviations ${\sigma[\,{\mathscr A}({\bf a},\,\lambda)\,]}$ and ${\sigma[\,{\mathscr B}({\bf b},\,\lambda)\,]}$:
\begin{equation}
\sigma[\,{\mathscr A}({\bf a},\,\lambda)\,]\;\sigma[\,{\mathscr B}({\bf b},\,\lambda)\,]\,=\,
-\,{\bf D}({\bf a}){\bf D}({\bf b})\,=\,(-I\cdot{\bf a})(+I\cdot{\bf b})
\,=\,{\bf a}{\bf b}\,=\,{\bf a}\cdot{\bf b}\,+\,{\bf a}\wedge{\bf b}\,.
\end{equation}
This product, however, is precisely the rotor that quantifies the twist in the fibration of the
3-sphere \ocite{Christian}\ocite{Chris-macro}.
Its value varies from ${+1}$ for ${{\bf b}={\bf a}}$ to ${-1}$ for
${{\bf b}=-\,{\bf a}}$ and back, producing the correct combination of observed probabilities \ocite{origins}.

\subsection{What can we learn from Weatherall's analysis?}

It should be evident from the above results that there is a valuable lesson to be learned from Weatherall's analysis:
An explicit, constructive, quantitatively precise physical model X cannot be undermined by repudiating its distorted
misrepresentation Y, even by appealing to a formal theorem (especially when that theorem is grounded on unphysical
assumptions). Such a strategy only serves to exemplify an elementary logical fallacy---namely, the straw-man fallacy.

\section{Another Explicit Event-by-Event Simulation of the Local-Realistic Model for the EPR-Bohm Correlation}

\begin{figure}[t]
\hrule
\vspace{-3.4cm}
\scalebox{1.74}{
\begin{pspicture}(-1.0,-0.05)(1.0,5.9)

\epsfig{figure=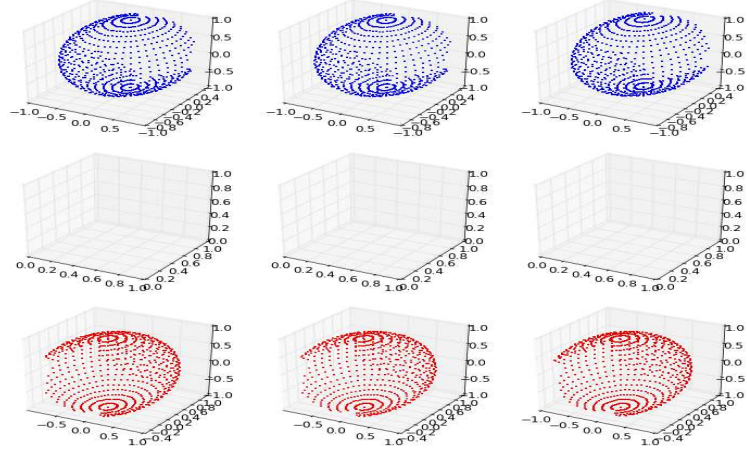,height=102pt,width=169.8pt}

\end{pspicture}}
\vspace{0.2cm}
\hrule
\caption{The correlations vanish (middle row) for ${\sin^2(\theta)=\text{constant}}$.
Rows from top to bottom show outcomes
${+1}$, ${0}$, and ${-1}$. Columns from left to right show Alice's, joint, and Bob's outcomes. 3D coordinates
are ${({\bf e}_o,\,{\bf x},\,|{\cal C}|)}$, with ${|{\cal C}|}$ being the radius.}
\label{fig-588778}
\vspace{0.3cm}
\hrule
\end{figure}

Inspired by the simulation of the 3-sphere model by Chantal Roth discussed in the previous appendix, Michel Fodje has
produced another explicit, event-by-event simulation of the model that is worth elaborating on here.
While the simulation by Chantal Roth is primarily
based on the joint probability density function ${|{\cal C}({\bf a},\,{\bf b};\,{\bf e}_o)|}$ defined in (\ref{c-23-want}),
the simulation by Michel Fodje is based on the individual probability density functions ${|{\cal C}({\bf a};\,{\bf e}_o)|}$ and
${|{\cal C}({\bf b};\,{\bf e}_o)|}$. From footnote 1 above we see that the individual probability
density functions ${p({\bf a};\,{\bf e}_o,\,\lambda)=|\cos(\,\eta_{{\bf a}\lambda{\bf e}_o})|}$ satisfy the relation
\begin{equation}
\frac{1}{4\pi} \int_{S^2} \,|\cos(\,\eta_{{\bf a}\lambda{\bf e}_o})|\;\,d\Omega\,
=\,\frac{1}{2}\,\sin^2(\,\eta_{{\bf a}{\bf e}_o})\bigg|_0^{\frac{\pi}{2}}=\,\frac{1}{2}
\end{equation}
with respect to any fixed vector ${\bf a}$. This suggests that the constraints
\begin{equation}
|\cos(\,\eta_{{\bf a}{\bf e}_o})|\,\geq\,
\frac{1}{2}\,\sin^2(\,\theta\,)\,\leq\,|\cos(\,\eta_{{\bf b}{\bf e}_o})| \label{con-michel}
\end{equation}
for arbitrary angles ${\eta_{{\bf a}{\bf e}_o}\in[0,\,2\pi)}$ and ${\eta_{{\bf b}{\bf e}_o}\in[0,\,2\pi)}$ should play a
crucial role in dictating the integrity and
strength of the correlation between the results ${{\mathscr A}({\bf a};\,{\bf e}_o,\,\theta)}$
and ${{\mathscr B}({\bf b};\,{\bf e}_o,\,\theta)}$, for any given angle ${\theta\in[0,\,\pi/2]}$.
According to these constraints the probability
densities ${|\cos(\,\eta_{{\bf a}{\bf e}_o})|}$ and ${|\cos(\,\eta_{{\bf b}{\bf e}_o})|}$ for observing the measurement
results ${{\mathscr A}({\bf a};\,{\bf e}_o,\,\theta)}$ and ${{\mathscr B}({\bf b};\,{\bf e}_o,\,\theta)}$ depend on
the common parameter ${\theta}$, just as they depend on the common random vector ${{\bf e}_o}$. This in turn suggests that
we may treat ${\theta}$ as an additional random parameter, and take the set
\begin{equation}
\left\{({\bf e}_o,\,\theta)\;\bigg|\;|\cos(\,\eta_{{\bf x}{\bf e}_o})|\,\geq\,
\frac{1}{2}\,\sin^2(\,\theta\,)\;\;\,\forall\;{\bf x}\in{\rm I\!R}^3\right\} \label{thesette}
\end{equation}
as the complete or initial state of our physical system. Given this complete state, the outcomes of measurements are
deterministically determined by the topological constraints within the 3-sphere. From a geometrical point of view, the
parameter ${\theta}$ links two disconnected ``sections'' of ${S^3}$ ({\it i.e.}, two ``orthogonal'' 2-spheres within
${S^3}$) defined by the bivectors ${I\cdot{\bf a}}$ and ${I\cdot{\bf b}}$, by means of the constraints (\ref{con-michel})
[see Fig.${\,}$\ref{fig-588778} for the relationship between these two 2-spheres].
 
\begin{figure}[t]
\hrule
\vspace{-3.4cm}
\scalebox{3.4}{
\begin{pspicture}(0.48,0.1)(5.0,4.65)

\epsfig{figure=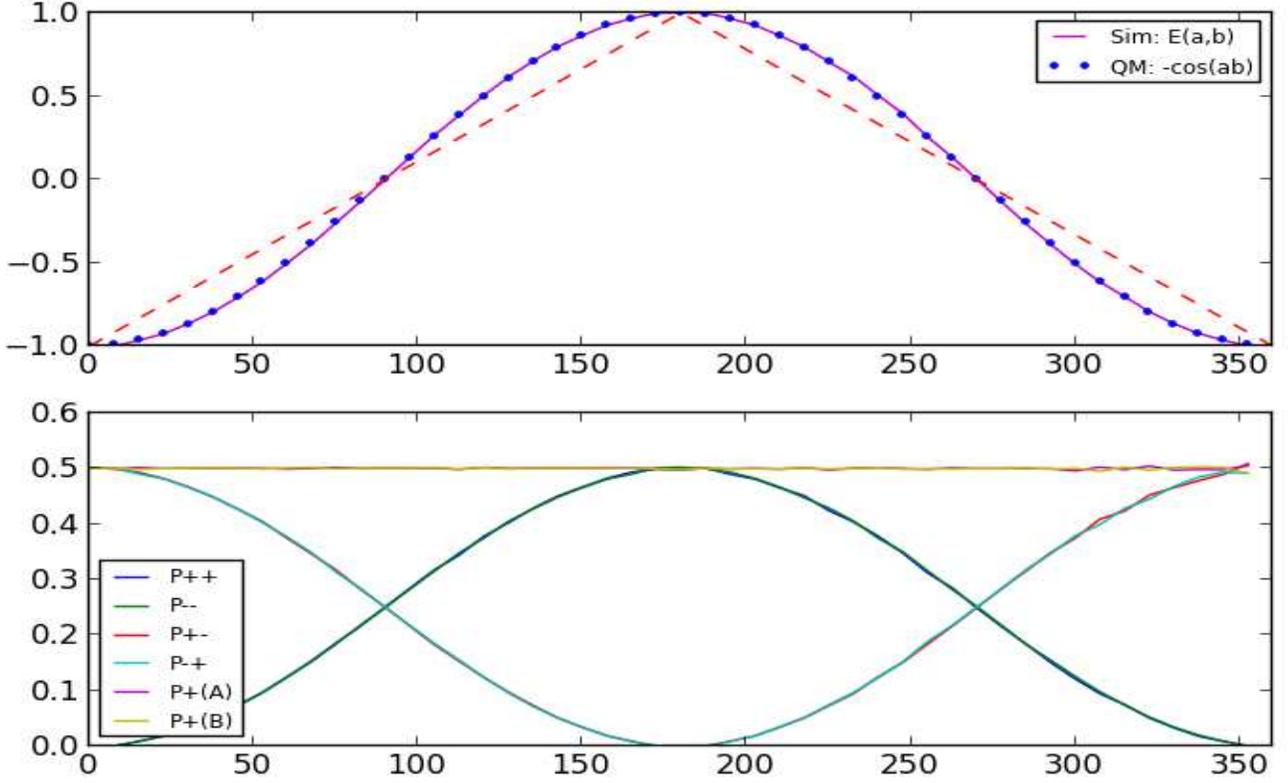,height=102pt,width=169.8pt}

\end{pspicture}}
\vspace{0.2cm}
\hrule
\caption{Another explicit, event-by-event simulation of the correlation between simultaneously occurring measurement events
${{\mathscr A}=\pm1}$ and ${{\mathscr B}=\pm1}$ within a parallelized 3-sphere. The code for this simulation
is written by Michel Fodje, in Python. Along with other relevant information, it can be downloaded form the
simulation's website https://github.com/minkwe/epr-simple/.}
\label{fig-5998}
\vspace{0.3cm}
\hrule
\end{figure}

With these considerations, we define the measurement functions simply as
\begin{align}
S^3\ni\pm\,1\,=\,{\mathscr A}({\bf a};\,{\bf e}_o,\,\theta)
\,&=\,sign\{\,-\cos(\,\eta_{{\bf a}{\bf e}_o}\,)\},\,\;\;\text{for a given}\,\;\theta\in[0,\,\pi/2]\,, \label{mmm999-oi}\\
\text{and}\;\;\;S^3\ni\pm\,1\,=\,{\mathscr B}({\bf b};\,{\bf e}_o,\,\theta)
\,&=\,sign\{\,+\cos(\,\eta_{{\bf b}{\bf e}_o}\,)\},\,\;\;\text{for a given}\,\;\theta\in[0,\,\pi/2]\,,\label{999mmm-oi}
\end{align}
where ${{\bf e}_o}$ and ${\theta}$ are common to both ${{\mathscr A}({\bf a};\,{\bf e}_o,\,\theta)}$ and
${{\mathscr B}({\bf b};\,{\bf e}_o,\,\theta)}$, and ${\bf a}$ and ${\bf b}$ are specific instances of the vector ${\bf x}$.

Once again let me stress the obvious that these functions define manifestly {\it local} measurement outcomes.
What is more, given the initial or complete state ${\{{\bf e}_o, \,\theta\}}$, the outcomes 
${{\mathscr A}({\bf a};\,{\bf e}_o,\,\theta)}$ and ${{\mathscr B}({\bf b};\,{\bf e}_o,\,\theta)}$ are
deterministically determined to be either ${+1}$ or ${-1}$, for any freely chosen vectors
${\bf a}$ and ${\bf b}$. Here ${{\bf e}_o}$ is a random vector on ${S^2}$
as defined before, and ${\theta}$ is a random angle, chosen from the interval ${[0,\,\pi/2]}$.
The correlation is then calculated quite simply as
\begin{align}
{\cal E}({\bf a},\,{\bf b})\,=\lim_{\,n\,\gg\,1}\left[\frac{1}{n}\sum_{i\,=\,1}^{n}\,
{\mathscr A}({\bf a},\,{\bf e}^i_o,\,\theta^i)\;{\mathscr B}({\bf b},\,{\bf e}^i_o,\,\theta^i)\right]
\,=\,-\,{\bf a}\cdot{\bf b}\,. \label{notafort}
\end{align}
 
What is more, for the measurement functions defined in (\ref{mmm999-oi}) and (\ref{999mmm-oi}) the probabilities of observing
the specific outcomes  ${+1}$ or ${-1}$ turn out to be exactly ${1/2}$, with 100\% detector efficiency. In other
words, every particle that emerges in a state ${({\bf e}_o, \,\theta)}$ defined by the set (\ref{thesette})
ends up being detected by the detector,
just as in the previous simulation. On the other hand, the probabilities of jointly observing the results
${{\mathscr A}({\bf a};\,{\bf e}_o,\,\theta)}$ and ${{\mathscr B}({\bf b};\,{\bf e}_o,\,\theta)}$
turn out to be exactly those predicted by quantum mechanics
[cf. Eqs.${\,}$(\ref{lesson}) to (\ref{lesson-6})]. Consequently, not only the correlations between these
results turn out to be those predicted by quantum mechanics [cf. Eq.${\,}$(\ref{molnot})],
but also the Clauser-Horne inequality is necessarily violated in this simulation,
just as it is violated in the previous simulation [cf. Eq.${\,}$(\ref{pp2})].

When we celebrate Bell's theorem next year for surviving for 50 years, let us hope we do not ignore these results. 

\vspace{-0.29cm}

\renewcommand{\bibnumfmt}[1]{\textrm{[#1]}}

\end{document}